\documentclass[%
 reprint,
superscriptaddress,
preprintnumbers,
nofootinbib,
 amsmath,amssymb,
 aps,
]{revtex4-1}

\usepackage[colorlinks]{hyperref}
\usepackage{graphicx}
\usepackage{dcolumn}
\usepackage{bm}
\usepackage{color}
\usepackage[dvipsnames]{xcolor}
\usepackage{placeins}
\usepackage{slashed}
\usepackage{multirow}
\usepackage{subfigure}
\usepackage{tabularx}
\usepackage{mathtools}
\usepackage{floatrow}
\usepackage{blindtext}
\usepackage{soul}
\usepackage{slashed}

\definecolor{coral}{RGB}{255,127,80}
\definecolor{indigo}{RGB}{75,0,130}
\definecolor{red}{rgb}{0.9, 0,0}
\definecolor{cerulean}{rgb}{0., 0.62,0.9}
\definecolor{navy}{rgb}{0.05, 0.05,0.8}

\newcommand{\mim}{m_{\slashed{E}}}

\newfloatcommand{capbtabbox}{table}[][\FBwidth]
\renewcommand{\eqref}[1]{Eq.~\ref{#1}}

\begin{document}

\title{Lepton-flavor violating axions at MEG II}

\author{Yongsoo Jho}
\affiliation{Department of Physics and IPAP, Yonsei University, Seoul 03722, Republic of Korea}
\affiliation{CERN, Theoretical Physics Department, Geneva, Switzerland}
\author{Simon Knapen}
\affiliation{Berkeley Center for Theoretical Physics, Department of Physics, University of California, Berkeley, CA 94720, USA}
\affiliation{Theoretical Physics Group, Lawrence Berkeley National Laboratory, Berkeley, CA 94720, USA}
\author{Diego Redigolo}
\affiliation{CERN, Theoretical Physics Department, Geneva, Switzerland}
\affiliation{INFN, Sezione di Firenze Via G. Sansone 1, 50019 Sesto Fiorentino, Italy}

\preprint{CERN-TH-2022-044}

\date{\today}

\begin{abstract}
We study the sensitivity of the existing MEG data to lepton flavor violating axion-like particles produced through \mbox{$\mu^+ \to e^+ a \gamma$} and estimate the discovery potential for the upcoming MEG~II experiment in this channel. The MEG~II signal efficiency can be improved significantly if a new trigger can be implemented in a dedicated run with a reduced beam intensity. This search would establish the world leading measurement in this channel with only 1 month of data taking. 
\end{abstract}

\maketitle

\section{Introduction}

Despite the advances in precision flavor measurements, the Standard Model (SM) flavor puzzle remains one of its greatest mysteries. The SM is equipped with three generations of fermions, which come with an elaborate set of flavor symmetries. These flavor symmetries are (weakly) broken by the SM yukawa couplings and the mechanism generating the neutrino masses. If this breaking occurs \emph{spontaneously}, one expects a set of pseudo-goldstone bosons with flavor violating couplings~\cite{Wilczek:1982rv,Reiss:1982sq,Gelmini:1982zz,Feng:1997tn}. Focusing on the leptonic sector, lepton flavor violating (LFV) axion-like particles (ALPs) can also arise in QCD axion models where the Peccei-Quinn symmetry is embedded non-trivially in the SM flavor group~\cite{Ema:2016ops,Calibbi:2016hwq,Linster:2018avp,Calibbi:2020jvd}, in familon models explaining the leptonic mass hierarchies \`a la Froggatt-Nielsen~\cite{Froggatt:1978nt,Calibbi:2020jvd} as well as in majoron models generating neutrino masses~\cite{Calibbi:2020jvd,Ibarra:2011xn,Garcia-Cely:2017oco,Heeck:2019guh}. In these constructions the ALP mass can be very light and its decay constant is typically very large, resulting in ALP lifetimes longer than the age of the Universe. This allows for the intriguing possibility that a LFV ALP can be the Dark Matter (DM).  

The large decay constant of the ALP suppresses its interactions with the SM, which makes it  challenging for any laboratory experiment to test it. However, the presence of LFV couplings provides a unique opportunity to probe  new physics at high scales through the exotic decays of SM particles to the light ALP. Here we study LFV ALPs in rare muon decays such as $\mu^+\to e^+ a$ and $\mu^+\to e^+ a\gamma$, where the stable or long-lived ALP $a$ remains invisible to the detectors.
Such rare muon decays can be tested at exquisite precision by the next generation muon experiments at the Paul Scherrer Institute (PSI) if dedicated data taking strategies are implemented. 

\begin{figure}[h]
\centering
\includegraphics[width=1\textwidth]{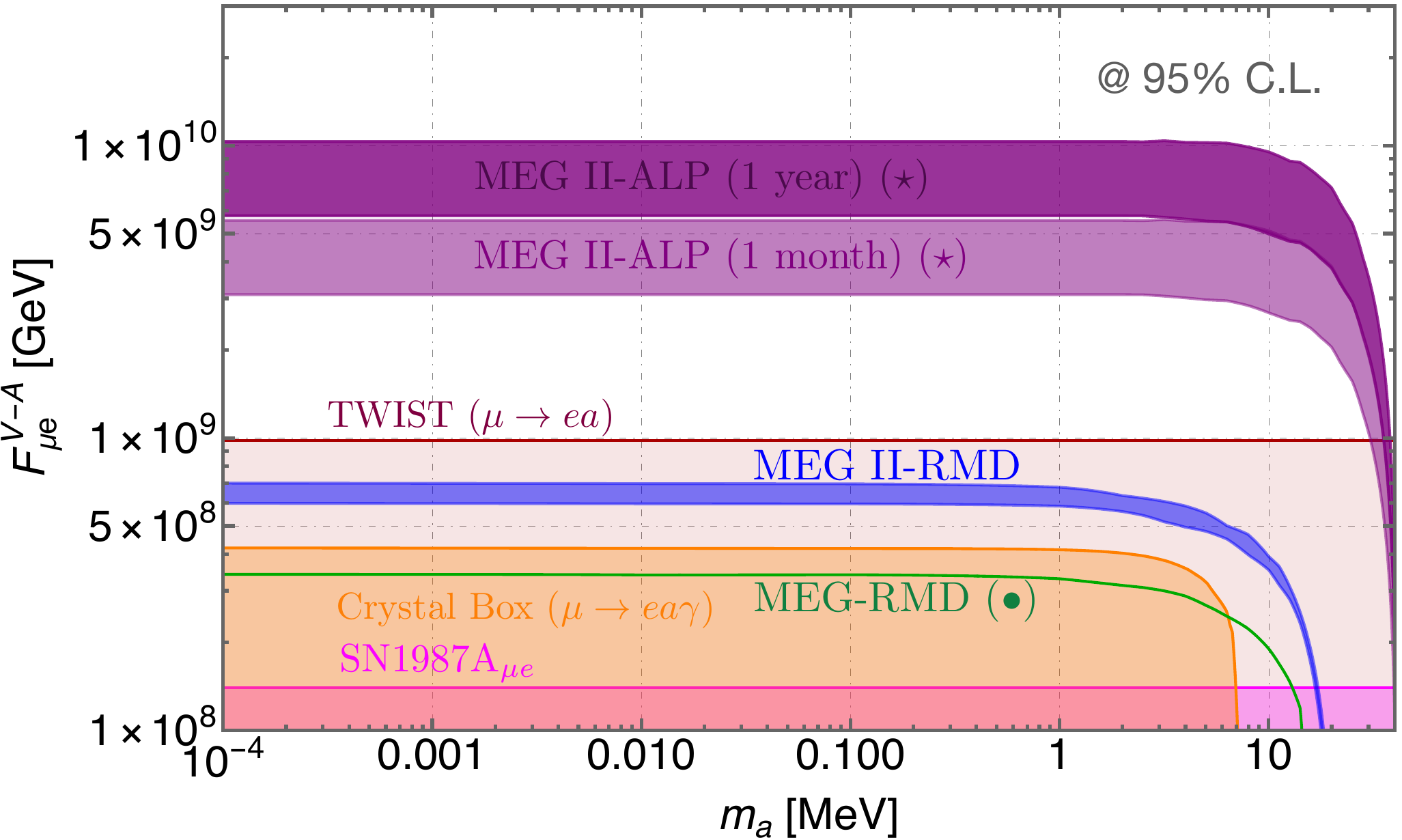}
\caption{ 95\% C.L. limits on $F_{\mu e}^{V-A}$. The {\bf green} line is the expected bound from the parasitic analysis of MEG RMD data~\cite{MEG:2013mmu} (Sec.~\ref{sec:MEGlimit}). The {\bf blue} band is the MEG II projection of the same parasitic analysis (Sec.~\ref{sec:MEGIIlimit}). The upper boundary of the band correspond to  a 50\%  reduction of the RC background with respect to the MEG search~\cite{MEG:2013mmu}. The {\bf purple} bands show the reach of the new hypothetical MEG II run with lower beam intensity and a dedicated trigger stream, with 1 month and 1 year of data taking (Sec.~\ref{sec:MEGII}). The upper (lower) limit of the reach corresponds to the lower (upper) limit in the determination of the trigger rate as detailed in Fig.~\ref{fig:trigger}. The {\bf orange} shaded region is the most conservative Crystal Box bound derived in \cite{Calibbi:2020jvd}. The {\bf dark red} shaded region is the bound from the TWIST experiment on $\mu^+\to e^+ a$~\cite{Bayes:2014lxz}. The {\bf magenta} shaded region is the supernova bound on the LFV coupling derived in~\cite{Calibbi:2020jvd}.}
\label{fig:money}
\end{figure}

The main objective of our study is to identify a new data-taking strategy for MEG II~\cite{MEGII:2018kmf} that maximizes its sensitivity to $\mu^+\to e^+ a\gamma$. Along the way, we  show that the existing MEG data~\cite{MEG:2013mmu} should already yield a competitive limit, though we lack some information to perform a faithful recast of the data. As shown in Fig.~\ref{fig:money}, this expected limit  competes with the current best bound set by the Crystal Box experiment~\cite{Bolton:1988af} for ALP masses larger than 8 MeV. 

Experimentally, the missing mass variable in the \mbox{$\mu^+\to e^+ a\gamma$} channel allows for a more robust background discrimination as compared to the \mbox{$\mu^+\to e^+ a$} channel. This is especially true for left-handed ALPs, for which the $\mu^+\to e^+ a$ channel gives a monocromatic line at the kinematic endpoint of the $\mu^+\to e^+ \nu_e\bar{\nu}_\mu$ background. This region is however typically assumed to be signal free and used for calibration purposes~\cite{Calibbi:2020jvd}. Accounting for the corresponding systematic uncertainties, the TWIST collaboration is setting the current best bound on left-handed LFV ALP couplings from $\mu^+\to e^+ a$~\cite{Bayes:2014lxz}.  

As shown in Fig.~\ref{fig:money}, a search for $\mu^+\to e^+ a\gamma$ at \mbox{MEG II} (in blue) can approach the current TWIST limit (dark red), but an experimental challenge remains: the existing triggers are optimized for MEG II's flagship analysis in the $\mu^+\to e^+ \gamma$ channel but have a suboptimal acceptance for $\mu^+\to e^+ a\gamma$. We explore an alternative data-taking strategy which greatly increases the signal acceptance by adjusting the trigger selection while reducing the beam intensity. This approach can improve on the TWIST limit with only one month of data taking as shown by the purple solid line in Fig.~\ref{fig:money}.  

The LFV ALP is defined by the low energy effective action
\begin{equation}
\label{couplings}
\mathcal{L}_{\rm{ eff}}^{\text{LFV}} \supset\frac{\partial_\mu a}{2 f_a} \, \bar \mu \gamma^\mu (C^V_{\mu e} + C^A_{\mu e} \gamma_5 ) e +\text{h.c.} \, , 
\end{equation}
where $C^{V}_{\mu e}$ ($C^{A}_{\mu e}$) controls the vector (axial) LFV coupling. For concreteness we focus in the main text on left-handed ALP couplings, setting $C^A_{\mu e}=-C^V_{\mu e}=C^{V-A}_{\mu e}$ and define the shorthand notation $F_{\mu e}^{V-A}=\sqrt{2} f_a/C^{V-A}_{\mu e}$. The cases of right-handed ALP couplings, with $C^A_{\mu e}=C^V_{\mu e}$, or purely axial (vectorial) with $C^V_{\mu e}=0$ ($C^A_{\mu e}=0$) will be discussed in Appendix~\ref{app:othercases} for completeness. The kinematical distributions and branching ratio for $\mu^+\to e^+ a \gamma$ were computed for a massless and a massive ALP~\cite{Hirsch:2009ee,Calibbi:2020jvd}, assuming an unpolarized muon. Here we further extend these results by accounting for the muon polarization, which is relevant for MEG~\cite{MEG:2015kvn}. The fully differential decay width for $\mu^+\to e^+ a\gamma$ is given in Appendix~\ref{app:signal}.

Our paper is organized as follows: in Sec.~\ref{sec:trigger} we review the standard MEG trigger selection, derive the expected MEG limit on $\mu^+\to e^+ a \gamma$  in Sec.~\ref{sec:MEGlimit} and a projection for MEG II in Sec.~\ref{sec:MEGIIlimit}. In Sec.~\ref{sec:MEGII} we explore an alternative data-taking strategy optimized for the ALP signal. We conclude in Sec.~\ref{sec:discussion} with a discussion of the physics potential for light new physics at muon facilities, as well as the theory motivation for searches of this class. In Appendix~\ref{app:signal} we detail our new signal computation. Appendix~\ref{app:validation} contains a validation of our simulation framework and more details on our analysis. In Appendix~\ref{app:othercases} we present the expected reach for different chiral structures of the ALP couplings.

\section{Existing and planned datasets}\label{sec:MEG}

In the MEG and MEG II experimental setup a high intensity $\mu^+$ beam is stopped in a thin target located at the center of a magnetic spectrometer. The main detectors making up the experiment are a high resolution liquid xenon scintillation detector and a drift chamber, optimized to measure the outgoing photon and positron respectively. The experiment is further equipped with a timing counter of scintillator bars at MEG and scintillator tiles at MEG II, to provide a good timing measurement for the $e^+$ and to aid with the trigger selection \cite{Adam:2013vqa}.

\subsection{The MEG trigger}
\label{sec:trigger}
We first describe the standard MEG trigger~\cite{Galli:2014uga}, which is now being upgraded with increased bandwidth but similar logic for MEG II~\cite{Francesconi:2019tnq,Galli:2019nmv}. The trigger is  optimized to look for the $\mu^+ \to e^+ \gamma$ decay, which amounts to requiring the positron and photon to be \emph{back-to-back} with energies $E_{e,\gamma}\simeq m_\mu/2$~\cite{MEG:2013oxv}. As a consequence, the trigger is suboptimal to probe $\mu^+ \to e^+ a \gamma$, where the signal rate is maximized for a soft photon, \emph{collinear} with the positron.  

\begin{figure*}[t]
\centering
\includegraphics[width=0.32\textwidth]{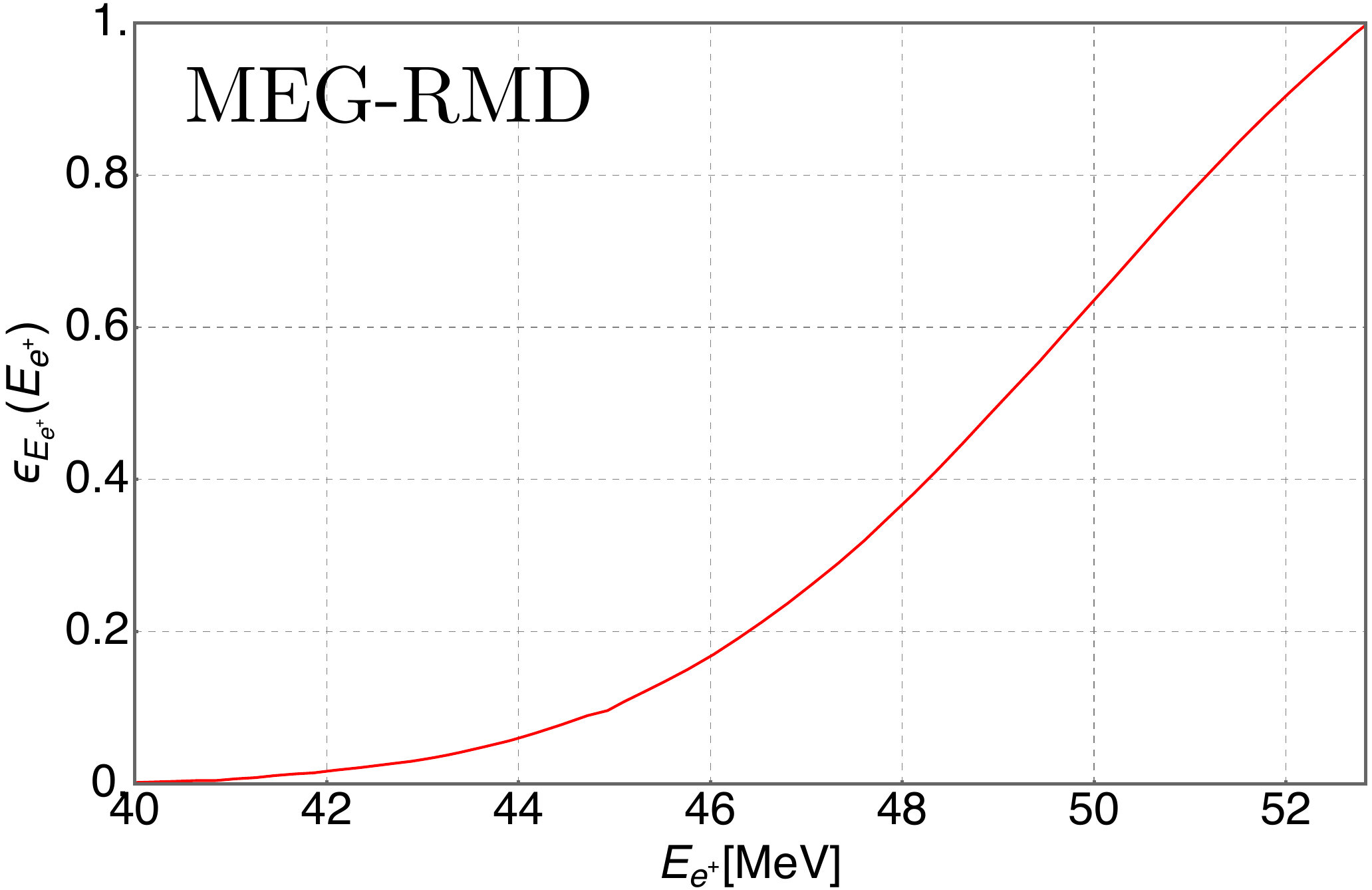}
\includegraphics[width=0.32\textwidth]{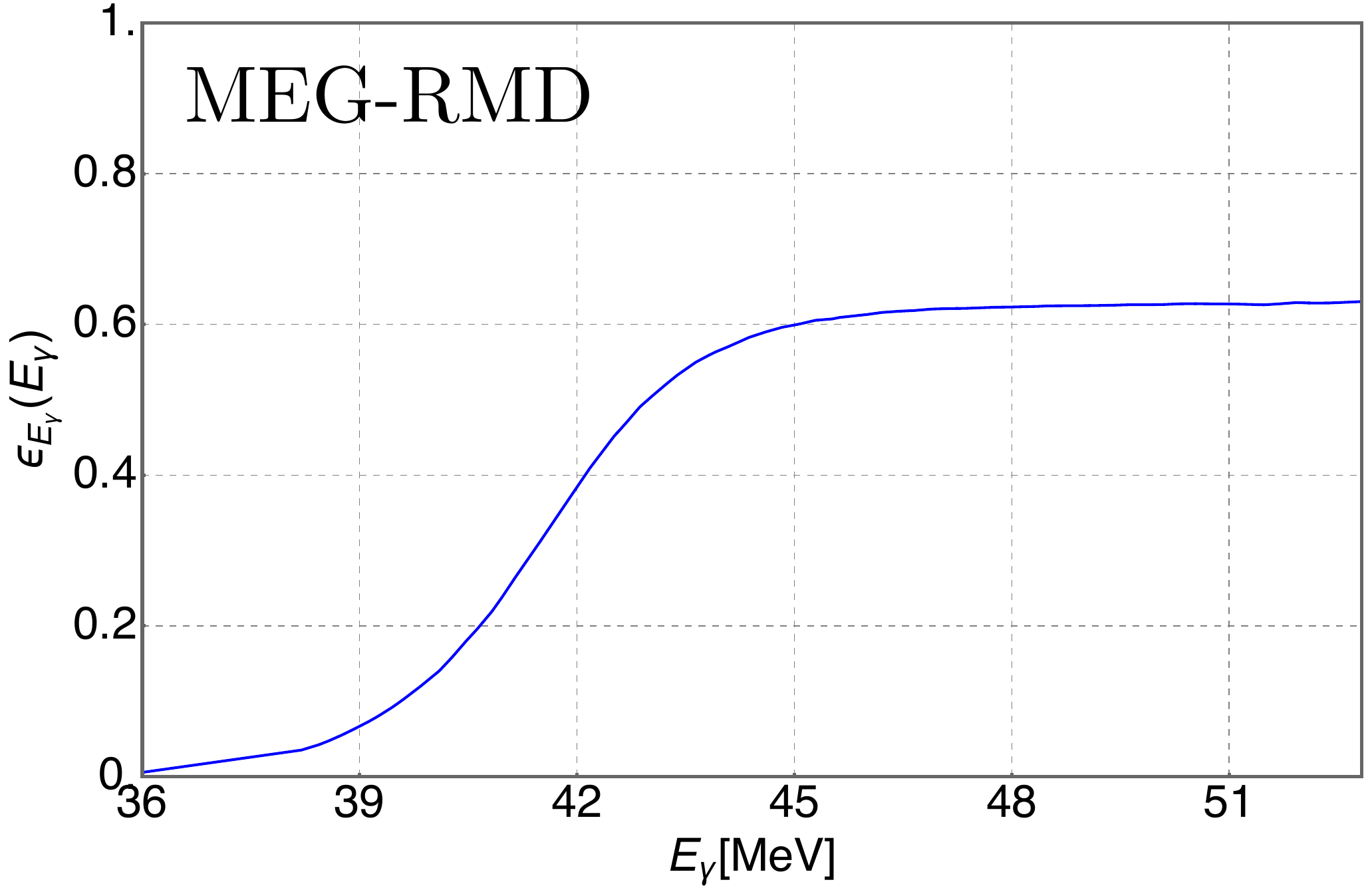}
\includegraphics[width=0.32\textwidth]{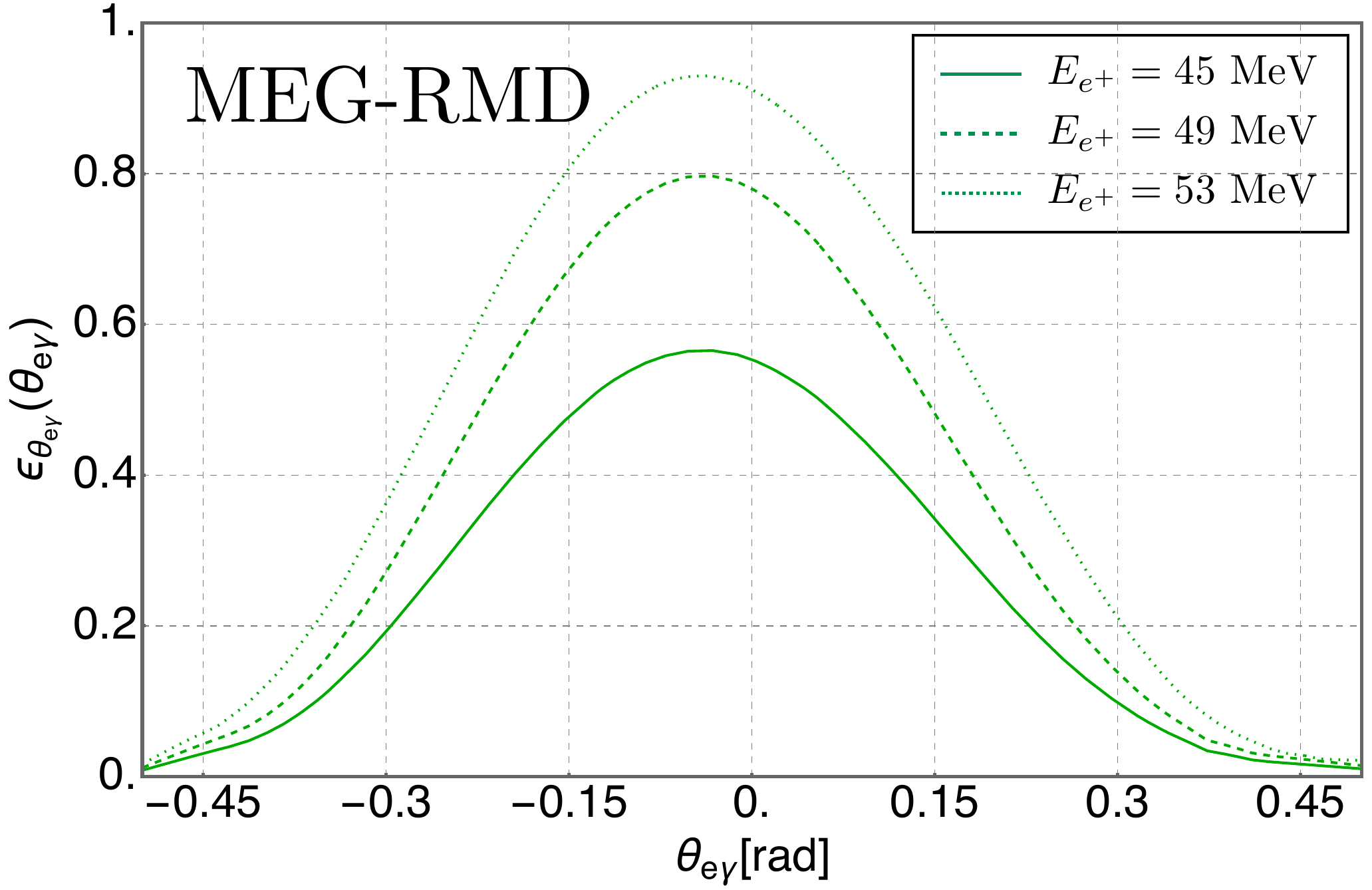}
\caption{{\bf Left:} Positron energy-dependent part $\epsilon_{E_{e}}$ of the MEG  trigger efficiency {\bf Middle:} Photon energy-dependent part $\epsilon_{E_\gamma}$ of the MEG trigger efficiency. {\bf Right:} polar angle-dependent part $\epsilon_{\theta_{e\gamma}}$ of the MEG trigger efficiency. The total trigger efficiency is given by Eq.~\ref{MEG_eff_approx} up to a normalization factor $c_{\text{RMD}}=0.35$, which is defined in \eqref{eq:norm} to reproduce the number of observed RMD events $N_{\text{RMD}}\vert_{\text{obs.}}=12900$ in Ref.~\cite{MEG:2013mmu}.}
\label{Fig:MEGrec}
\end{figure*}

At trigger level, the only available information is the photon energy, the time and the conversion point measured by liquid xenon scintillation detector and the hit and time measured by the timing counter~\cite{Adam:2013vqa}. Because of the positron spectrometer design, requiring a hit in the timing counter corresponds to selecting positrons with energies higher than roughly 45 MeV. In addition, an extra trigger selection on the photon energy of $E_\gamma\gtrsim 40\text{ MeV}$ is imposed to keep the trigger rate below $10\text{ Hz}$, as required by the experimental design. The positron (photon) energy trigger efficiency $\epsilon_{E_{e}}$ ($\epsilon_{E_{\gamma}}$) is a function of the positron (photon) energy only $E_{e}$ ($E_\gamma$) as long as the they are within the detector acceptance. $\epsilon_{E_{e}}$ ($\epsilon_{E_{\gamma}}$) is plotted in the left (central) panel of Fig.~\ref{Fig:MEGrec}, as taken from Ref.~\cite{MEG:2013mmu}. 

The information of the full positron momentum as measured by the drift chamber cannot be accessed at trigger level~\cite{Galli:2014uga}. In the standard MEG trigger algorithm, the coordinates of the positron hit in the timing counter are matched to the muon stopping point by assuming that the positron momentum and direction are consistent with those of a $\mu^+ \to e^+ \gamma$ decay.  The trigger therefore selects predominantly back-to-back positron-photon pairs. The dependence of the trigger efficiency on the polar angle between the positron and the photon ($\theta_{e\gamma}$) depends on the energy of the positron, while the dependence on the azimuthal angle ($\phi_{e\gamma}$) is a subdominant effect after the trigger energy cuts on positron and photons are imposed.\footnote{Following the MEG notation, we define $\theta_{e\gamma} \equiv \pi - \theta_e - \theta_\gamma$ and $\phi_{e\gamma} = \pi + \phi_e - \phi_\gamma$ as the polar and azimuthal angle in between the positron and the photon, respectively. In this notation the back-to-back topology corresponds to $\theta_{e\gamma} = \phi_{e\gamma} = 0$.} In the right panel of Fig.~\ref{Fig:MEGrec} we show the trigger efficiency $\epsilon_{\theta_{e\gamma}}$ as a function of $\theta_{e\gamma}$ for different values of the photon energy $E_\gamma$. As expected, the closer the photon energy is to $m_{\mu}/2$, the more efficient the trigger is in the region of $\theta_{e\gamma}\simeq0$. 

We use the MEG measurement of the radiative muon decay (RMD) $\mu^+ \to e^+ \nu \bar{\nu} \gamma$~\cite{MEG:2013mmu} to obtain quantitative information about the MEG trigger. The search is based on $N_{\mu^+,\text{tot}}^{\rm MEG}=1.8 \times 10^{14}$ muons collected in the years \mbox{2009 - 2010} with a beam intensity of \mbox{$R_{\mu^+}^{\text{MEG}}=3\times10^7 \mu^+/\text{sec}$}. The MEG collaboration measures the turn-on of the trigger efficiency relative to a prescaled trigger with a lower threshold, and obtains the overall normalization from their (internal) Monte Carlo simulation. The full, differential trigger efficiency as a function of $E_{e}$, $E_{\gamma}$, $\phi_{e\gamma}$ and $\theta_{e\gamma}$ was not made public and we must therefore construct  an approximate model from the published turn-on curves in Fig.~\ref{Fig:MEGrec}. We do so by assuming that the full efficiency function factorizes as
\begin{equation}
\epsilon_{\rm trigger}^{\rm MEG}  \equiv \epsilon_{E_{e}}(E_{e}) \times \epsilon_{E_{\gamma}}(E_{\gamma}) \times \epsilon_{\theta_{e\gamma}}(E_{e}, \theta_{e\gamma})\ , \label{MEG_eff_approx}
\end{equation}
and by extrapolating the functional dependence of $\epsilon_{\theta_{e\gamma}}$ as
\begin{equation}
\epsilon_{\theta_{e\gamma}}\!(E_{e}, \theta_{e\gamma})\!=\!\left[-2.6 + \frac{0.07 E_{e}}{\text{ MeV}}\right]\epsilon_{\theta_{e\gamma}}(49\,\text{MeV},\theta_{e\gamma})\, . 
\end{equation}
For the geometric acceptance of the photon detector we take  $\theta_\gamma \in [70^\circ, 110^\circ]$, $\phi_\gamma \in [-60^\circ, +60^\circ]$ \cite{MEGII:2018kmf}. The positron timing is  detector not hermetic but was designed to detect  $E_e=m_\mu/2$ positrons that are back-to-back to the photons that are within the acceptance of the calorimeter. We therefore estimate its acceptance to be $\phi_e \in [120^\circ, 240^\circ]$. Due the non-homogeneous magnetic field, $\phi_e$ acceptance interval should shift for lower values of $E_e$, but we cannot reliably model this effect without the full MEG simulation framework. 

With this procedure, we reproduce all kinematical distributions in Ref.~\cite{MEG:2013mmu} up to an overall normalization factor, as we show in Appendix~\ref{app:validation}. This offset of the overall rate between the data and our simulations could be due to the simplifying assumptions above or other more subtle experimental effects, either in the trigger or in the offline selection. In addition to the acceptance cuts described above, we further assume that the offline positron acceptance in $\theta_{e\gamma}$ is the same as the trigger acceptance, shown in the right-hand panel of Fig.~\ref{Fig:MEGrec}, which is likely an overestimate. We therefore introduce an overall normalization factor, $c_{\text{RMD}}$, to rescale our simulations such that they match the number of observed RMD events after the offline kinematic selection:
\begin{equation}
\begin{split}
&E_e > 45 \text{ MeV}\ ,\quad \ E_\gamma > 40 \text{ MeV}\, , \\
& |\theta_{e\gamma}|< 0.3 \ ,\quad  |\phi_{e\gamma}|< 0.3 \ . \label{MEG_kin}
\end{split}
\end{equation}
With the available information we cannot unambiguously attribute  $c_{\text{RMD}}$ to our modeling of either the trigger or the off-line selection, which will be a source of uncertainty when we estimate the trigger rate later in this section. Concretely, $c_{\text{RMD}}$ is defined as
\begin{equation}
c_{\text{RMD}}\equiv\frac{N_{\text{RMD}}\vert_{\text{obs.}}}{N_{\mu^+,\text{tot}}^{\rm MEG}\cdot \text{BR}_{\text{RMD}}^{\text{base}}\cdot\epsilon^{\rm trig.}_{\text{RMD}} \cdot   \epsilon^{\rm off.}_{\text{RMD}}/\epsilon^{\rm trig.}_{\text{RMD}} }\ , \label{eq:norm}
\end{equation} 
and we find it to be $c_{\text{RMD}}\simeq 0.35$. The inputs to \eqref{eq:norm} were found as follows: $N_{\text{RMD}}\vert_{\text{obs.}}=12900$ is the observed number of RMD events in Ref.~\cite{MEG:2013mmu}.  To ensure it is finite, the RMD branching ratio was defined subject to an arbitrary, minimal set of baseline cuts.\footnote{Our baseline cuts are $E_{e}>40\text{ MeV}$, $E_{\gamma}>5\text{ MeV}$.} The offline angular acceptance of the positron was taken to be the same as the trigger acceptance, in the right-hand panel of Fig.~\ref{Fig:MEGrec}. The muon polarization was taken to be $P_\mu = -0.85$, as measured in MEG~\cite{MEG:2015kvn}.  which give  $\text{BR}_{\text{RMD}}^{\text{base}}=1.44\times 10^{-5}$ with the formula in Refs.~\cite{Fronsdal:1959zzb,Kuno:1999jp}. The analysis is not sensitive to these baseline cuts, as long as they are looser than the trigger cuts.
 Starting from this baseline branching ratio, we can use our Monte Carlo to compute the online efficiency  $\epsilon^{\rm trig.}_{\text{RMD}}$ by applying \eqref{MEG_eff_approx}, and the offline efficiency $\epsilon^{\rm off.}_{\text{RMD}}$ by applying both \eqref{MEG_eff_approx} and  \eqref{MEG_kin}. This yields $\epsilon^{\rm trig.}_{\text{RMD}}= 3.90 \times 10^{-5}$ and $\epsilon^{\rm off.}_{\text{RMD}}/\epsilon^{\rm trig.}_{\text{RMD}} =0.36$, which serve as inputs for \eqref{eq:norm}.  
 
The MEG trigger selects RMD events together with random coincidences (RC), which are generated when a photon from an RMD $\mu^+ \to (e^+) \nu \bar{\nu} \gamma$ (with a missing soft positron) and an positron from an unrelated Michel decay \mbox{$\mu^+ \to e^+ \nu \bar{\nu}$} are detected as coming from the same event.  These pileup events are due to the enormous intensity of the muon beam, which is only partially offset by the strict cuts on the time separation between the positron and the photon. The RC background also receives a contribution from positrons annihilating in flight into a pair of photons, when one of the two photons is lost and the other is paired up with a hard positron from the Michel decay. This positron annihilation contribution is not explicitly included in our simulation but we can roughly account for it by normalizing the total RC measured offline to $N_{\rm RC}\vert_{\text{obs.}}=83850$, which is the number of RC MEG observed after their offline selection cuts~\cite{MEG:2013mmu}. Analogously to the RMD discussion, we can write  
\begin{equation}
N_{\mu^+,\text{tot}}^{\rm MEG}\cdot\text{BR}^{\text{base}}_{\text{RC}}\cdot\epsilon^{\text{trig.}}_{\text{RC}}\cdot \epsilon^{\text{off.}}_{\text{RC}}/\epsilon^{\text{trig.}}_{\text{RC}}=N_{\rm RC}\vert_{\text{obs.}}\ , \label{eq:RCfix}
\end{equation}
where $\text{BR}^{\text{base}}_{\text{RC}}$ is the probability of a muon to be involved in an RC event. In this sense it can be thought of as the baseline ``branching ratio'' of the random coincidences and it is defined as  
\begin{equation}
\text{BR}^{\text{base}}_{\text{RC}}=c_{\text{RC}}\cdot\text{BR}^{\text{base}'}_{\text{RMD}}\cdot \text{BR}^{\text{base}}_{\text{Mich.}}\cdot R_{\mu^+}\cdot \Delta t_{e\gamma}^{\text{trig.}} ,\label{defNRC}
\end{equation}
where $\Delta t_{e\gamma}^{\text{trig.}}\simeq24$ ns is the trigger resolution on the arrival time between the measured photon and positron ~\cite{Galli:2014uga}. The $c_{\text{RC}}$ parameter is the overall normalization constant we use to normalize our Monte Carlo to the MEG data and is the RC analogue of the $c_{\text{RMD}}$ parameter in \eqref{eq:norm}. It is fixed from \eqref{eq:RCfix} and \eqref{defNRC}. $\text{BR}^{\text{base}'}_{\text{RMD}}$ is obtained with our Monte Carlo and is defined by requiring the positron to be outside the detector acceptance or softer than 40 MeV, and the photon to have $E_{\gamma}>5\text{ MeV}$ and be within the geometrical acceptance of the detector. The resulting value is $\text{BR}^{\text{base}'}_{\text{RMD}}=2.50 \times 10^{-3}$, while $\text{BR}^{\text{base}}_{\text{Mich.}}=0.28$ is the branching ratio of the Michel decay $\mu^+\to e^+\nu\bar{\nu}$ after the minimal energy cut $E_e>40\text{ MeV}$ and the geometrical acceptance are applied. The baseline RC differential distributions are then obtained by assuming RMD photons and Michel positron to be time coincident. This simplification should capture the kinematic properties of the main component of the RC background.  Analogous to the RMD background,  $\epsilon^{\rm trig.}_{\text{RC}}$ is found with our Monte Carlo by applying \eqref{MEG_eff_approx} and the offline efficiency $\epsilon^{\rm off.}_{\text{RC}}$ is obtained by applying both \eqref{MEG_eff_approx} and  \eqref{MEG_kin}. This yields $\epsilon^{\rm trig.}_{\text{RC}}=6.82 \times 10^{-4}$  and $\epsilon^{\rm off.}_{\text{RC}}/\epsilon^{\rm trig.}_{\text{RC}} =0.02$, which serve as inputs for \eqref{eq:RCfix}. With these inputs we find $c_{\text{RC}}=0.07$.

For purposes that will be clear in Sec.~\ref{sec:MEGII}, we here estimate the trigger rate of both the RMD and the RC events at MEG by computing the total number of simulated events passing the trigger selection and dividing the effective run time, which we take to be \mbox{$t_{\text{run}}^{\text{MEG}}=N_{\mu^+,\text{tot}}^{\rm MEG}/R_{\mu^+}^{\text{MEG}}=6\times 10^6\text{ sec}$}. 
When doing so, we must account for the fact that the online timing window is $\Delta t_{\gamma e}^{\text{trig.}}\simeq24 \text{ ns}$~\cite{Galli:2014uga}, roughly 6 times larger than the offline window $\Delta t_{\gamma e}^{\text{off.}} = 4 \text{ ns}$. This increases the RC trigger rate with a factor of $\Delta t_{\gamma e}^{\text{trig.}}/ \Delta t_{\gamma e}^{\text{off.}}\simeq6$. 
A large uncertainty on our estimate comes from the overall normalization of our efficiencies $c_{\text{RMD}}$ and $c_{\text{RC}}$ (see \eqref{eq:norm} and \eqref{defNRC})  as we cannot unambiguously determine whether our modeling of the online or offline selection is responsible for these correction factors.  In practice, our estimate of the RMD trigger rate can therefore vary within a factor of $1/c_{\text{RMD}}$ and the RC trigger rate within a factor of $1/c_{\text{RC}}$:  
\begin{align*}
&R_{\text{RMD}}^{\text{trig.}}\in \frac{N_{\text{RMD}}\vert_{\text{obs.}}}{ \epsilon^{\rm off.}_{\text{RMD}}\cdot t_{\text{run}}^{\text{MEG}}}\!\left(1,\frac{1}{c_{\text{RMD}}}\right)\!= (1.7-4.8)\!\cdot\!10^{-2}\text{ Hz}\,,\\
&R_{\text{RC}}^{\text{trig.}}\in \frac{N_{\text{RC}}\vert_{\text{obs.}}}{ \epsilon^{\rm off.}_{\text{RC}}\cdot t_{\text{run}}^{\text{MEG}}}\!\left(1,\frac{1}{c_{\text{RC}}}\right)\!= (0.7-10)\text{ Hz}\, .
\end{align*}
Our estimated trigger rate is thus in the $1$-$10$ Hz range and completely dominated by the RC, for which the rate at trigger level is roughly a factor of 200 larger than the RMD rate. In Tab.~\ref{table:benchmark} and Fig.~\ref{fig:trigger} we will account for this uncertainty when optimizing the selection for the dedicated $\mu^+\to e^+ \gamma a$ analysis.  The corresponding uncertainty on the reach is indicated by the purple bands in Fig.~\ref{fig:money}. We emphasize that this uncertainty in our projection is due to the uncertainty in our modeling of the MEG experimental setup; a full analysis by the MEG collaboration would not be subject to it.

 \begin{figure*}[t]
\centering
\includegraphics[width=0.49\textwidth]{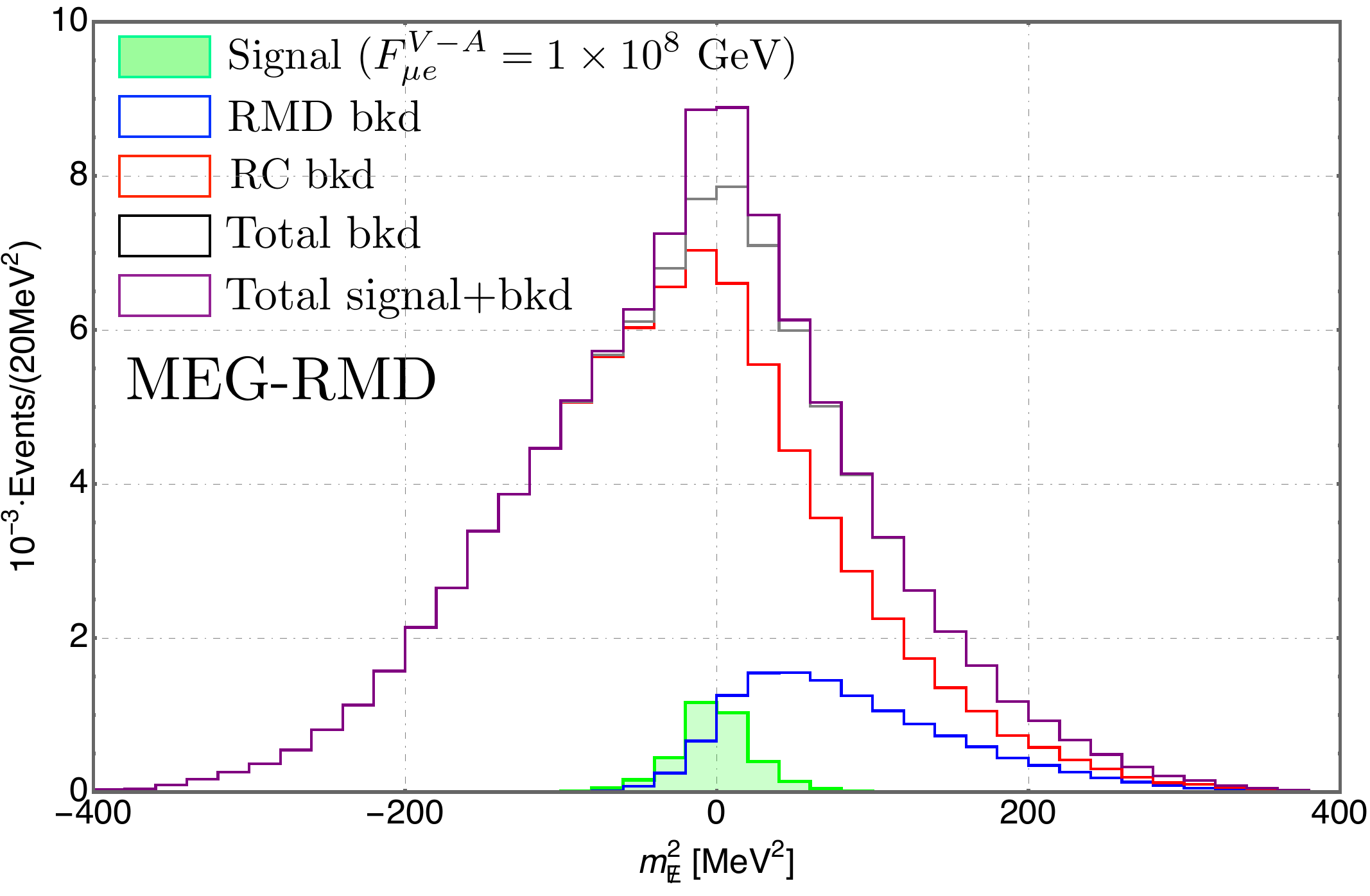}\hfill
\includegraphics[width=0.49\textwidth]{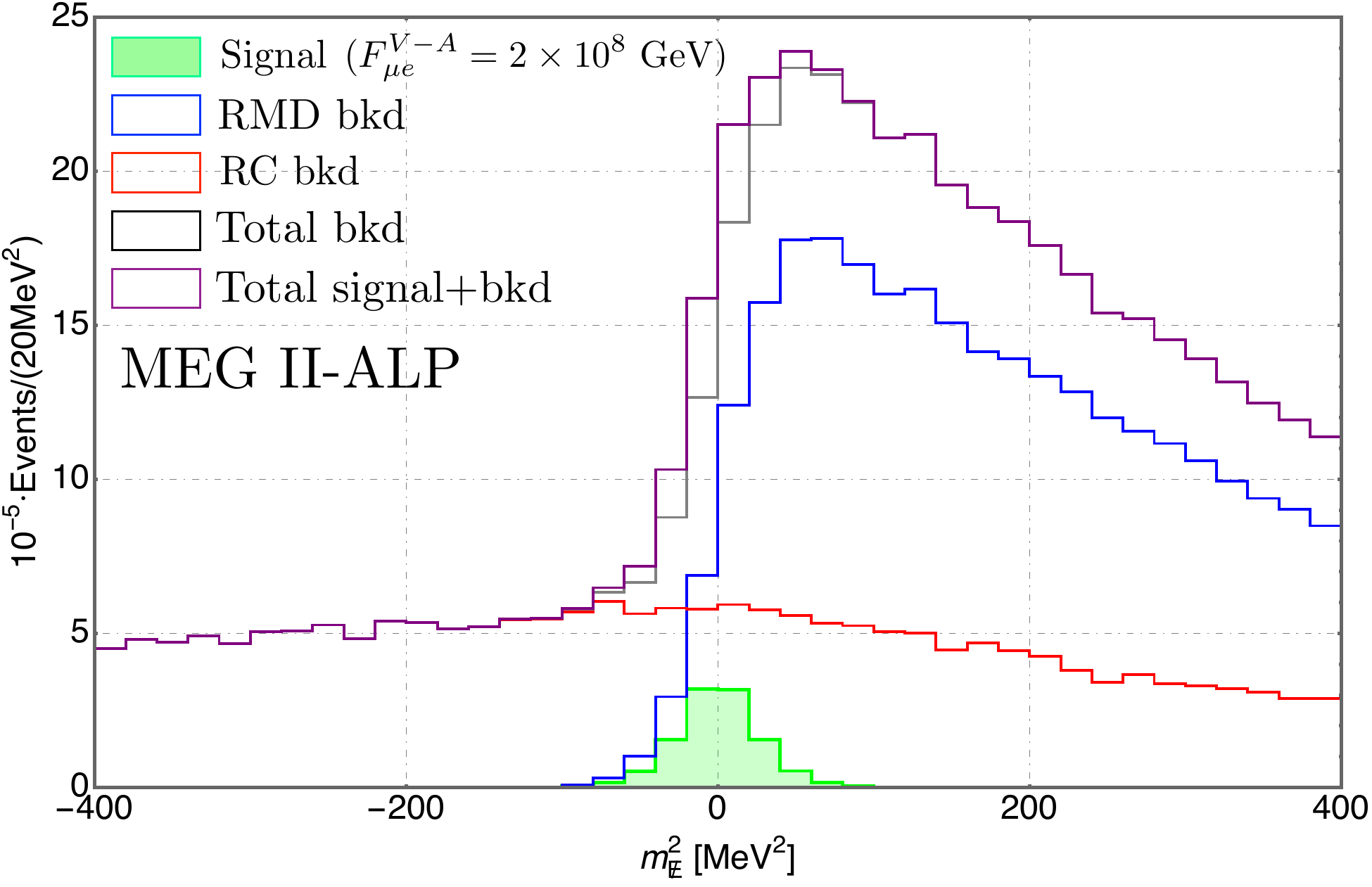}
\caption{Missing invariant mass distribution of signal and backgrounds after kinematic selection, detector efficiency and acceptance and resolution effects are taken into account. The ALP is fixed to $m_a = 10^{-4}$ MeV, which is effectively massless within the $m_{\slashed{E}}$ resolution.  {\bf Left:} missing mass distribution at MEG. {\bf Right:} Missing invariant mass distribution with the MEG II-ALP data taking strategy proposed in Sec.~\ref{sec:MEGII}. 
}
\label{fig:MMdist}
\end{figure*}
\begin{table*}[t]
\centering
\begin{tabular}{l|c|c|c|c|c|c}
\multirow{ 2}{*}{Search scenarios} & \multirow{ 2}{*}{$N_{\mu^+}^{\rm total}$} & \multirow{ 2}{*}{$\mathcal{R}_{\mu^+}$ [$\mu^+$/s]} & Trigger & \multirow{ 2}{*}{Trigger rate (Hz)} & Optimized & \multirow{ 2}{*}{$F_{\mu e}^{V-A}$ @ 95\% C.L.} \\
&&& selection & & $\Delta_{\rm \mim^2}$ & \\
\hline
MEG-RMD, {\color{Green} $\bullet$} & \ $1.8\times 10^{14}$ \ & $3\times 10^7$ & \eqref{MEG_eff_approx} & $\sim 0.77-10.37$ & $27$ MeV$^2$ & $3.4 \times 10^8$ GeV \\
\hline
MEG II-RMD & $1.8\times 10^{15}$ & $7\times 10^7$ & \eqref{MEG_eff_approx} & $\sim 0.21 - 25.19$ & $14$ MeV$^2$ & $(6-7)\times10^8$ GeV \\
\hline
MEG II-ALP, {\color{Purple} $\star$} & $1.8\times 10^{14}$ & ($1.6$-$5.8$)$\times 10^6$ & \eqref{MEGII_Eg_trig}+$\epsilon_{E_{e}}(E_{e})$ & $10$ & $35$ MeV$^2$ & $(5.8 - 10.3)\times 10^9$ GeV 
\end{tabular}
\caption{Summary of the searches for $\mu^+\to e^+ a \gamma$ at MEG and MEG II. The markers ({\color{Green} $\bullet$},{\color{Purple} $\star$}) correspond to the lines in Fig.~\ref{fig:money}. The expected limits are given for an effective massless axion (i.e.~with mass below the experimental resolution). MEG II-RMD limits vary depending on the normalization of the RC background, which can be reduced by 50\% w.r.t.~the measured value at MEG~\cite{MEG:2013mmu}. The trigger rate is estimated in more detail at the end of Sec.~\ref{sec:trigger}, with a factor of $1/c_{\text{RC}}\approx 14$ uncertainty. The latter affects the reach of the dedicated MEG II-ALP data taking run, where we fix the trigger rate to be 10 Hz and derive two optimal benchmark choices for the beam intensity. These different beam intensities each result in a slightly different projected bound, shown by the width of the purple bands in Fig.~\ref{fig:money}.}
\label{table:benchmark}
\end{table*}

\subsection{Parasitic analysis: expected MEG bound}\label{sec:MEGlimit}

We now show how the RMD measurement~\cite{MEG:2013mmu} can be repurposed as a search for $\mu^+\to e^+ a \gamma$. Concretely, we take the offline kinematic selection to be that in \eqref{MEG_kin}, which should be applied together with the trigger efficiency and the angular acceptances of the MEG detector: $\theta_\gamma \in [70^\circ, 110^\circ]$, $\phi_\gamma \in [-60^\circ, +60^\circ]$ and $\phi_e \in [120^\circ, 240^\circ]$. In the previous section, we explained how the factor $c_{\text{RMD}}$ is used to correct for our imperfect modeling of the detector efficiency for the RMD process. We assume that the same correction factor holds for the ALP signal. The missing mass ($\mim$) is defined as 
\begin{equation}
\mim^2 \equiv (m_\mu - E_e - E_\gamma)^2 - || \vec{p}_e + \vec{p}_\gamma ||^2\ .\label{eq:missing}
\end{equation}
The signal is a peak in the $\mim$ distribution, located at the ALP mass. The differential distributions of the signal, the RMD and the RC backgrounds are shown in the left-hand panel of Fig.~\ref{fig:MMdist}. 

The  final sensitivity depends on the energy and angular resolutions. For electron and photon energies between 40 and 53 MeV, the MEG detector resolutions are extracted from Ref.~\cite{MEG:2020zxk}, fitted and extrapolated to the energy range of interest. (See Appendix~\ref{app:validation}). From this procedure we derive the minimal resolution on the missing mass to be $4.5\text{ MeV}$. Any ALP with mass below this resolution will be seen as effectively massless by MEG. 

Assuming no bump in missing mass spectrum has been detected in the existing MEG data, we can estimate the expected limit with the following scheme: We take the signal ($\mathcal{S}$) and the background ($\mathcal{B} $) in a narrow $\mim^2$ window, where the window size, $\Delta_{\rm \mim^2}=27\text{ MeV}^2$, is chosen to optimize the sensitivity under the assumption of negligible systematics. To further improve the sensitivity, we use a double-sided binned log-likelihood ratio on the $(\theta_{e\gamma},\phi_{e\gamma})$ distribution of the events passing the $\mim^2$ cut 
\begin{eqnarray}
\Lambda (S) & = & - 2\sum_i  \ln \frac{L_i(S_i)}{L_i (\hat{S}=0)}\ ,
\end{eqnarray}
where $S_i$ ($B_i$) is the number of signal (background) events in $i$th bin of a grid with binsize 20 mrad $\times$ 20 mrad. The likelihood is defined as the poisson distribution
\begin{equation}
L_i (S_i) = \frac{(S_i + B_i)^{B_i}}{B_i !}  e^{-(S_i+B_i)},
\end{equation}
where we estimated the number of observed events in each bin with the expectation value of the background, $B_i$. Demanding $\Lambda (S) < 4$, we obtain the 95\% confidence level projected limit on $F_{\mu e}^{V-A}$, as shown in Fig.~\ref{fig:money}.

The projected bound from MEG data is slightly weaker than the most conservative bound from Crystal Box derived in Ref.~\cite{Calibbi:2020jvd} for an effectively massless ALP.\footnote{The Crystal Box collaboration gives a bound on the measured branching ratio $\text{Br}(\mu^+\to e^+a\gamma)<1.1\times 10^{-9}$ at 90\% C.L.~\cite{Bolton:1988af} with measured energies $E_{\gamma,e^+}>38\text{ MeV}$ and $\theta_{e\gamma}< 0.7$. Translating this to the theory prediction is subject to a large uncertainty from the energy loss of the positron before reaching the detector. This was estimated to be at most 5 MeV by the collaboration.  The most conservative theory bound is then obtained assuming a truth-level positron energy cut of 43 MeV.} This is due to the larger angular acceptance of Crystal Box which compensates for its smaller luminosity (\mbox{$N_{\mu^+}^{\text{Crystal Box}}=8\times 10^{11}$}) and its worse detector resolution.

\subsection{Parasitic analysis: MEG II projection}\label{sec:MEGIIlimit}

We now look into the future, assessing the MEG II projected sensitivity on $\mu^{+}\to e^{+} a \gamma$. We consider the MEG kinematical selection in \eqref{MEG_kin} and derive the expected reach at MEG II accounting for \emph{i)} the larger luminosity, which we take to be $N_{\mu^+}^{\text{ MEGII}}=1.8\times 10^{15}$,  \emph{ii)} the improved offline energy and angular resolution. As detailed in Appendix~\ref{app:validation}, we rescale the MEG resolutions using the resolution information at $E_{e,\gamma}\simeq m_\mu/2$~\cite{MEGII:2018kmf} by assuming that the energy dependence is the same as at MEG. We also account for the expected suppression of the RC background due to the installation of the radiative decay counter to reject the soft positron in the forward direction at MEG II~\cite{MEGII:2018kmf}. 
The projected limit is shown by the blue band in Fig.~\ref{fig:money}, where the upper edge corresponds to a $50\%$ suppression of the RC. Despite the expected MEG II improvements, the kinematical selection of \eqref{MEG_kin} can likely not push the reach beyond the present TWIST bound, motivating the exploration of a new, optimized data taking strategy. 

\section{A dedicated run}\label{sec:MEGII}

\begin{figure}[h]
\centering
\includegraphics[width=1\textwidth]{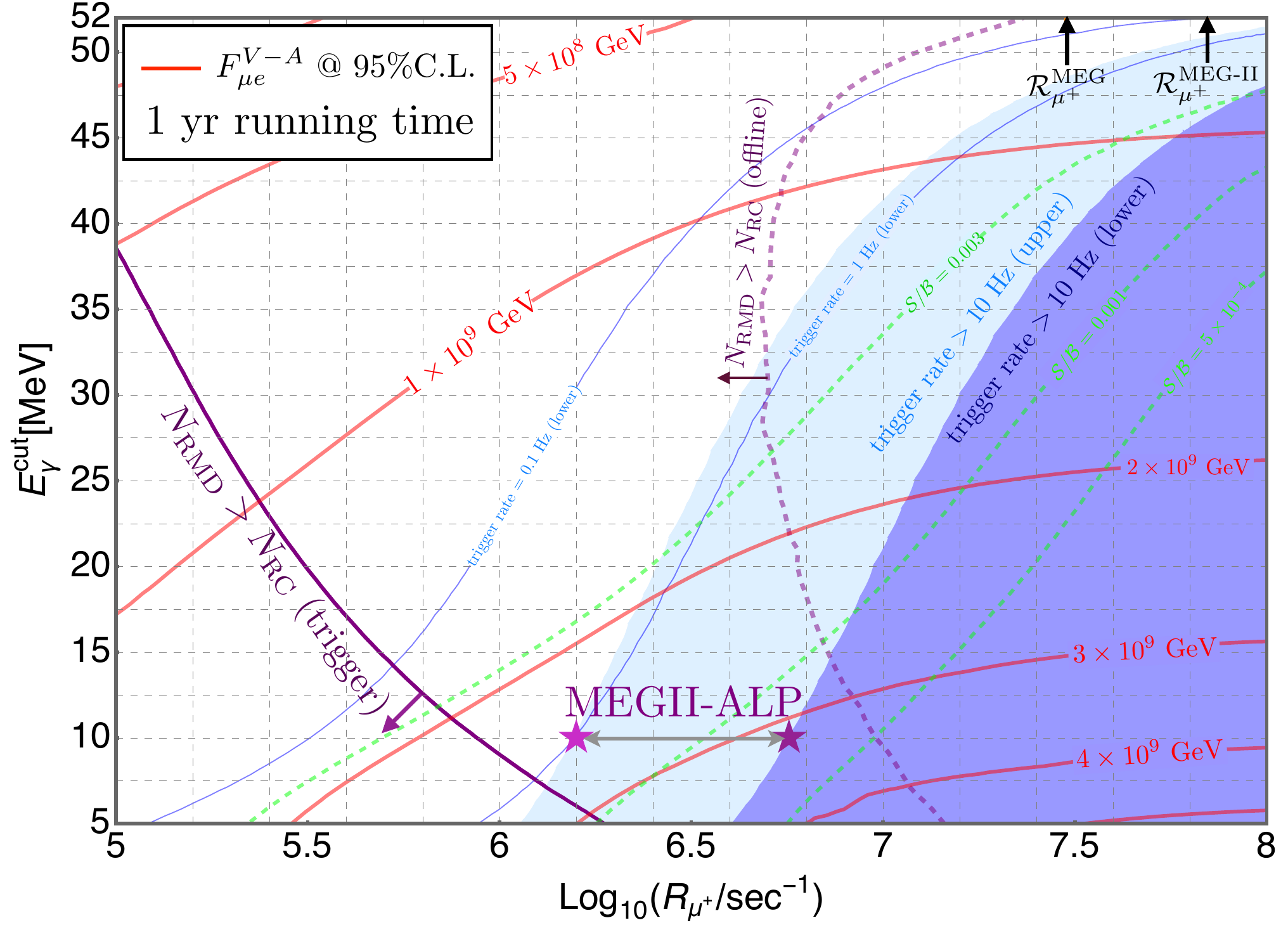}
\caption{MEG II-ALP performances as a function of the photon cut $E_\gamma^{\text{cut}}$ and of the beam intensity $R_{\mu^+}$. In {\bf red} contours of the 95\% C.L. reach after enforcing the kinematic selection $|m_{\slashed{E}}^2 - m_a^2 | < \Delta_{\rm \mim^2}$ and assuming 1 year of running time. The ALP is assumed to be massless within the experimental resolution. The {\bf blue} lines indicate 0.1 Hz and 1 Hz of the trigger rate, which gets larger than 10 Hz in the shaded regions. The {\bf dashed green} show contours of $\mathcal{S}/\mathcal{B}$. On the right of the {\bf solid purple} line the RMD background dominates over RC at trigger level, the {\bf dashed purple} line shows where the RMD dominates with the offline selection. The stars indicates the benchmarks chosen for the MEGII-ALP dedicated run, see \eqref{MEGII_Eg_trig}.}
\label{fig:trigger}
\end{figure}

To enhance the reach for $\mu^+ \to e^+ a \gamma$, one would ideally want to relax the energy and angular cuts on the photons while keeping the trigger rate below 10 Hz. This can be achieved by reducing the muon beam intensity $R_{\mu^+}$, which has the double benefit of \emph{i)} allowing the photon trigger cut to be looser, enhancing the signal acceptance and \emph{ii)} suppressing the RC background (which scales with $\sim R_{\mu^+}^2$) compared to the RMD background (which scales with $\sim R_{\mu^+}$).\footnote{We thank Luca Galli for suggesting to reduce the beam intensity.} In the remainder of this section we will estimate the sensitivity of such a hypothetical ``MEG II-ALP'' dedicated run. 

We define the experimental efficiency and acceptance by taking into account the turn-on of the positron trigger and the detector geometry only, which are defined as before. The detection efficiency as a function of $E_\gamma$, $\theta_{e\gamma}$ and $\phi_{e\gamma}$ are otherwise assumed to be one. This might be an optimistic assumption, which  can only be assessed by the MEG II collaboration.

In Fig.~\ref{fig:trigger} we study the signal and background acceptance as a function of the beam intensity $R_{\mu^+}$ and the lower bound on the photon energy $E_{\gamma}^{\text{cut}}$. For concreteness, we benchmark a trigger selection with 
\begin{equation}\label{MEGII_Eg_trig}
E_\gamma>10 \text{ MeV},\quad  R_\mu^+\lesssim (1.6-5.8)\times 10^6\, \mu^+/\text{sec}\, ,
\end{equation}
where the uncertainty on the optimal $R_\mu^+$ stems from our approximate estimate of the trigger rate in Sec.~\ref{sec:trigger} ($\star$ symbols in Fig.~\ref{fig:trigger}). The proposed data taking strategy requires the beam intensity to be reduced by roughly an order of magnitude compared to the MEG run, in order to keep the trigger rate below 10 Hz (see Table~\ref{table:benchmark}).  As can be seen from the purple line Fig.~\ref{fig:trigger}, lowering the photon energy cut together with the beam intensity makes the RMD background almost of the same order as the RC background, at trigger level. Loosening the photon energy cut as much as possible moreover maximizes the reach for the ALP signal. We expect the bottleneck of this strategy to be the energy threshold of the liquid scintillator, but at this time there is no public information about its response to low energy photons. For the purpose of our study we therefore select photon energies larger than $10\text{ MeV}$, where the detector efficiency should be excellent. The possibility of including softer photons can be considered by the MEG II collaboration. 

Offline, analogously to the previous section, we optimize the missing mass window to separate the signal from the background. The differential distributions are shown in Fig~\ref{fig:MMdist} right. We also perform the log-likelihood ratio test for the ($\theta_{e\gamma},\phi_{e\gamma}$) distribution to maximize the sensitivity. The detailed distribution of signal and background in the angular variables are given in Appendix~\ref{app:validation}. The optimal value for the width of the missing mass window is $\Delta_{\rm \mim^2}=35\text{ MeV}^2$, in the limit of negligible systematic uncertainty on the background. The broadening of the signal distribution can be traced back to the expected deterioration of the energy resolution on the photons at low energies, which is accounted by our fitting function of the resolution in Appendix~\ref{app:validation}.

The expected reach of this dedicated run is shown in Fig.~\ref{fig:money} for the same total luminosity as the MEG run $N_{\mu^+}=1.8\times 10^{14}$, which can be collected in a dedicated $1$ year run time ($\sim$ 50 weeks data taking) at the end of the commissioned run of MEG II. Interestingly, we show in Fig.~\ref{fig:money} that with only  $1$ month of data taking our proposal can already get the best sensitivity on left-handed LFV axions. Our projections neglect systematic uncertainties which can be parametrized in the cut and count scheme as $\mathcal{S}/\sqrt{\mathcal{B}+\eta_{\text{sys}}^2 (\mathcal{B}+\mathcal{S})^2}$. The contours of $\mathcal{S}/\mathcal{B}$ in Fig.~\ref{fig:trigger} indicate that the parameter $\eta_{\text{sys}}$ should be kept below $0.1\%$ in order for systematics uncertainties to be negligible. This assumption can again only be validated by the MEG collaboration.

\section{Discussion}\label{sec:discussion}

\begin{figure}[h]
\centering
\includegraphics[width=1\textwidth]{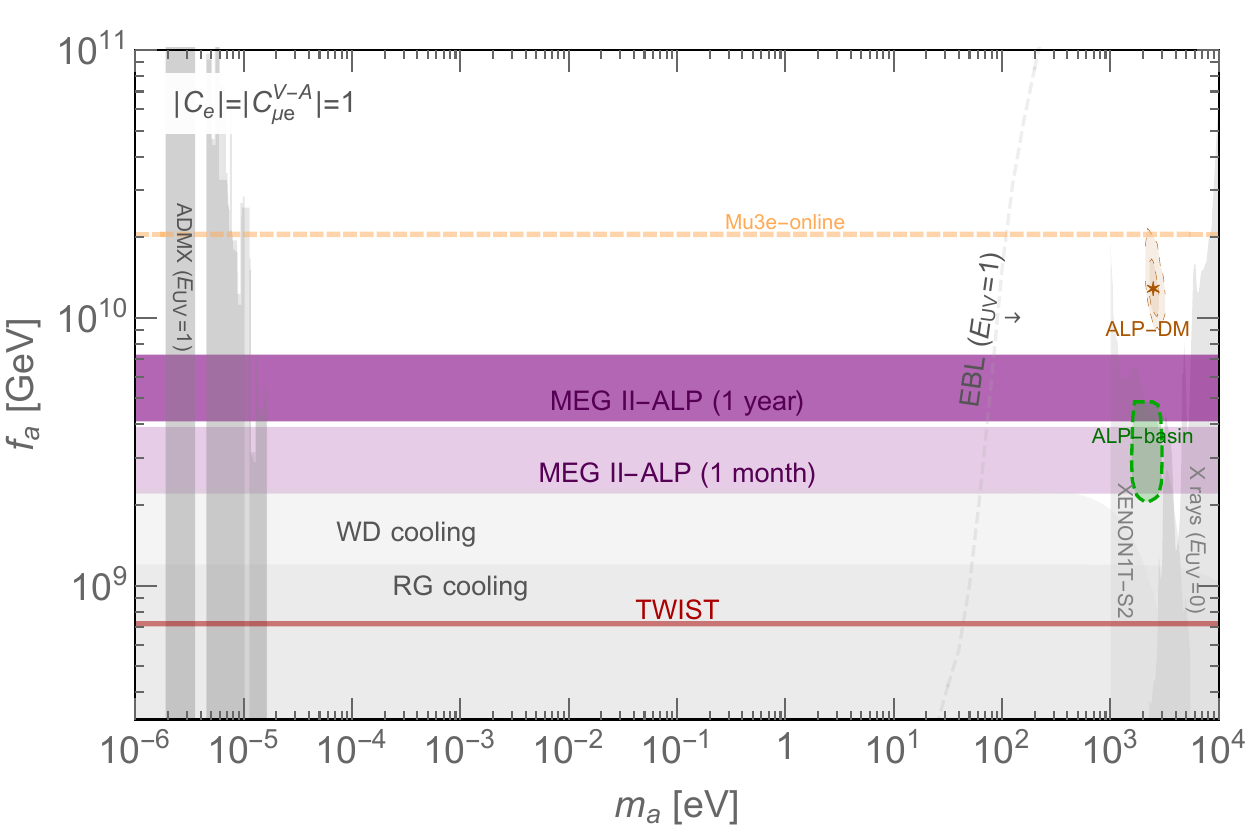}
\caption{ALP parameter space as a function of the decay constant $f_a$ and the mass $m_a$, assuming $C_{e}=C^{V-A}_{e\mu}=1$ and $E_{\text{UV}}=0$. The {\bf dark red} line is the present TWIST bound~\cite{Bayes:2014lxz}, while the {\bf purple} bands correspond the projections for the MEG II-ALP dedicated run shown in Fig.~\ref{fig:money}.  The solid/dashed line corresponds to 1 month/1 year of data taking. The {\bf dashed orange} line shows the (speculative) projection for a Mu3e online analysis of $\mu^+\to e^+ a$ data~\cite{Perrevoort:2018okj}. The {\bf shaded grey} regions show existing bounds from white dwarf (WD) and red giants (RG) cooling~\cite{Raffelt:1994ry,Viaux:2013lha,MillerBertolami:2014rka}, X rays searches of $\gamma$ lines from decaying DM~\cite{Boyarsky:2006hr,XQC:2015mwy}, absorption in direct detection experiments~\cite{PandaX:2017ock,XENON:2019gfn}, and existing resonant cavities~\cite{ADMX:2018ogs,ADMX:2018gho,ADMX:2019uok} for $E_{\text{UV}}=1$. The {\bf dashed grey} line show the bound on decaying DM from diffuse extra-galactic light observations~\cite{Creque-Sarbinowski:2018ebl} if $E_{\text{UV}}=1$ (the arrow points towards the excluded region). In the {\bf dark orange} blob ALP DM can explain the Xenon1T excess in electron recoils~\cite{XENON:2020rca,Takahashi:2020bpq,Bloch:2020uzh}, while in the {\bf dark green} region the solar basin can fit the same excess~\cite{VanTilburg:2020jvl}.}
\label{fig:theory}
\end{figure}
The experimental program for rare muon decays  has primarily focused on well motivated but very specific LFV final states such as $\mu^+\to e^+\gamma$ and $\mu^+\to e^+ e^-e^-$, with no (or very little) missing energy. These final states are very interesting tests of heavy new physics generating LFV operators of dimension six in the SM and can explore the flavor structure at the multi-TeV scale, for instance in supersymmetric or composite Higgs models (see for example Ref.~\cite{Calibbi:2017uvl}). They are however by design insensitive to signatures of low energy remnants of high scale LFV, such as light LFV axions.

The implementation of new trigger strategies can address this blind spot, by directly targeting events containing missing energy. These searches would enlarge the physics case of the muon experimental program in a completely orthogonal direction by testing dimension five operators with new, light long-lived particles that are very weakly coupled to the SM. In this context, rare muon decays can test scales as high as $10^{10}\text{ GeV}$ and probe non-trivial embeddings of the Peccei-Quinn symmetry inside the SM flavor group, as well as spontaneously broken lepton flavor symmetries more generally.  

An example in this direction is the online trigger strategy for $\mu^+\to e^+ a$ at the Mu3e experiment proposed in Ref.~\cite{Perrevoort:2018okj}, or the MEG II-fwd proposal put forward in Ref.~\cite{Calibbi:2020jvd}. Both these proposals are complementary to the one explored here, because they are expected to have limited sensitivity for a left-handed massless ALP: In particular, the whole MEG~II-fwd proposal ceases to be advantageous because the signal acceptance of left-handed ALPs is tiny in the forward region. The proposed search for Mu3e (orange dashed line in Fig.~\ref{fig:theory}) on the other hand faces severe challenges related to systematics uncertainties in hunting for a bump on top of the Michel end point. (This region is typically assumed to be signal-free and used for experimental calibration.) In addition, the MEG II experiment is already commissioned and should be able to perform the measurement on a shorter time scale than Mu3e. In the same spirit, we show in Appendix~\ref{app:othercases} the reach of our proposal on right-handed and vectorial/axial ALP couplings. With 1 year of data taking  MEG II can sensibly do better than the current best bound from the experiment performed by Jodidio et al. in 1986~\cite{Jodidio:1986mz} and set a bound which is only slightly weaker than the projections of Mu3e and MEG~II-fwd.

In Fig.~\ref{fig:theory} we show the impact of our projections in the ALP parameter space, assuming the flavor diagonal (FD) couplings to electrons 
\begin{equation}\label{eq:Ceedefinition}
\mathcal{L}_{\rm{ eff}}^{\text{FD}} \supset C_e\frac{\partial_\mu a}{2 f_a} \, \bar e \gamma^\mu  \gamma_5  e +\text{h.c.} \, , 
\end{equation}
are of the same order of the LFV coupling.\footnote{The $\partial_\mu a \, \bar e \gamma^\mu  e $ vanishes due to current conservation, up to a contribution to the $SU(2)_L$ anomaly.}  The coupling to photons 
\begin{equation}
 \mathcal{L}_{\rm{ eff}}^{\gamma\gamma}=\frac{E_{\text{eff}}}{f_a} \frac{\alpha}{8\pi} a F\tilde F
\end{equation}
 is controlled by $E_{\text{eff}}=E_{\text{UV}}+ C_{e} B(\tau_e)$, where $E_{UV}$ is the electromagnetic anomaly coefficient in the ultraviolet theory and $B(\tau)=\tau\,\text{arctan}^2(1/\sqrt{\tau-1})-1$ with $\tau_e=4 m_e^2/m_a^2-i\epsilon$ is the IR contribution from the electron threshold. We see that a \mbox{MEG II-ALP} dedicated run can probe new parameter space beyond the stellar cooling constraints already with 1 month of running.

A particularly interesting model is the  photophobic ALP with $E_{\text{UV}}=0$, which can be the DM with a mass $m_a\simeq 2-3 \text{ keV}$ and explain the recent XENON1T excess in electron recoils~\cite{XENON:2020rca,Takahashi:2020bpq,Bloch:2020uzh}, without being in tension with astrophysical bounds on decaying DM~\cite{Boyarsky:2006hr,XQC:2015mwy}. The same model could explain the Xenon excess if an ALP solar basin is formed around the Sun~\cite{VanTilburg:2020jvl} in a region of parameter space that is compatible with stellar energy losses ~\cite{Giannotti:2017hny}. Intriguingly, $E_{\text{UV}}=0$ is naturally realized in Majoron models where $C_{e}\sim C_{\mu e}^{V-A}\sim 1/16\pi^2$ are also generated after the right handed neutrinos are integrated out~\cite{Calibbi:2020jvd,Ibarra:2011xn,Garcia-Cely:2017oco,Heeck:2019guh}. From Fig.~\ref{fig:theory} we see that 1 year of running of MEG II-ALP will be sufficient to  probe the stellar basin explanation if $C_{e}\sim C_{\mu e}^{V-A}$.  

In conclusion, we hope that this study can pave the way for a more systematic assessment of the capabilities of MEG II in exploring light new physics with flavor violating couplings to the SM. In a first step, the existing and future data sets used for the RMD analysis can be (re)analyzed to obtain competitive limits on the $\mu^+\to e^+ a \gamma$ process. Second, a dedicated run of the MEG II experiment at lower beam intensity should yield a sensitivity surpassing the existing bounds by one order of magnitude. This program has the potential to shed light on open questions in axion phenomenology and even establish a new connection between precision measurements of muon branching ratios and ultralight DM candidates.

\section*{Acknowledgments}
We thank Marco Francesconi, Luca Galli, Angela Papa and Giovanni Signorelli for discussions about the MEG II detector. We thank Robert Ziegler and Lorenzo Calibbi for feedback on the draft. We especially thank Luca Galli for his extremely valuable feedback on the MEG trigger and his comments on the draft.  SK was supported by the Office of High Energy Physics of the U.S. Department of Energy under contract DE-AC02-05CH11231. YJ is supported in part by CERN-CKC graduate student fellowship program.

\appendix

\begin{figure*}[t]
\centering
\includegraphics[width=.32\textwidth]{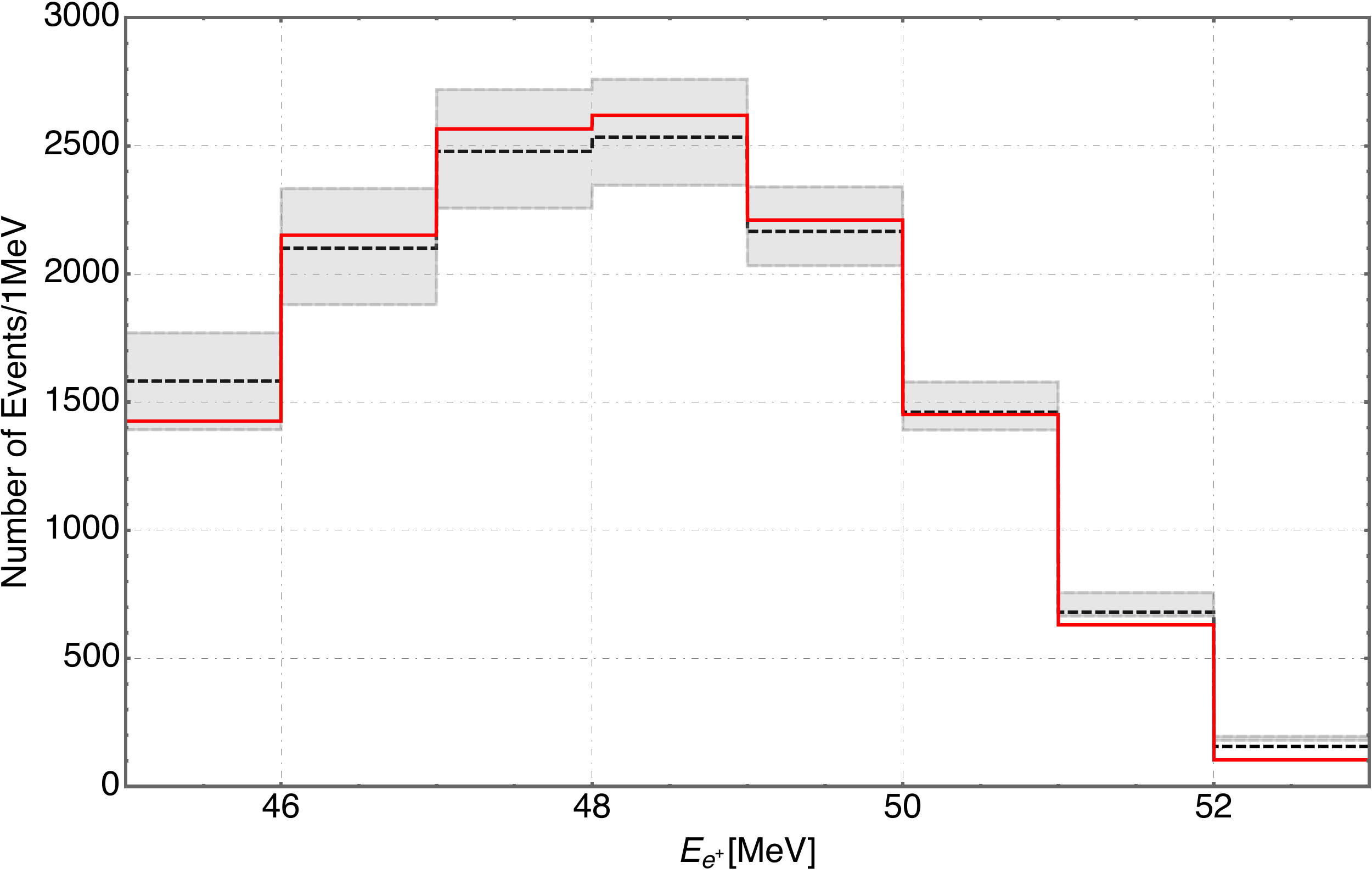}
\includegraphics[width=.32\textwidth]{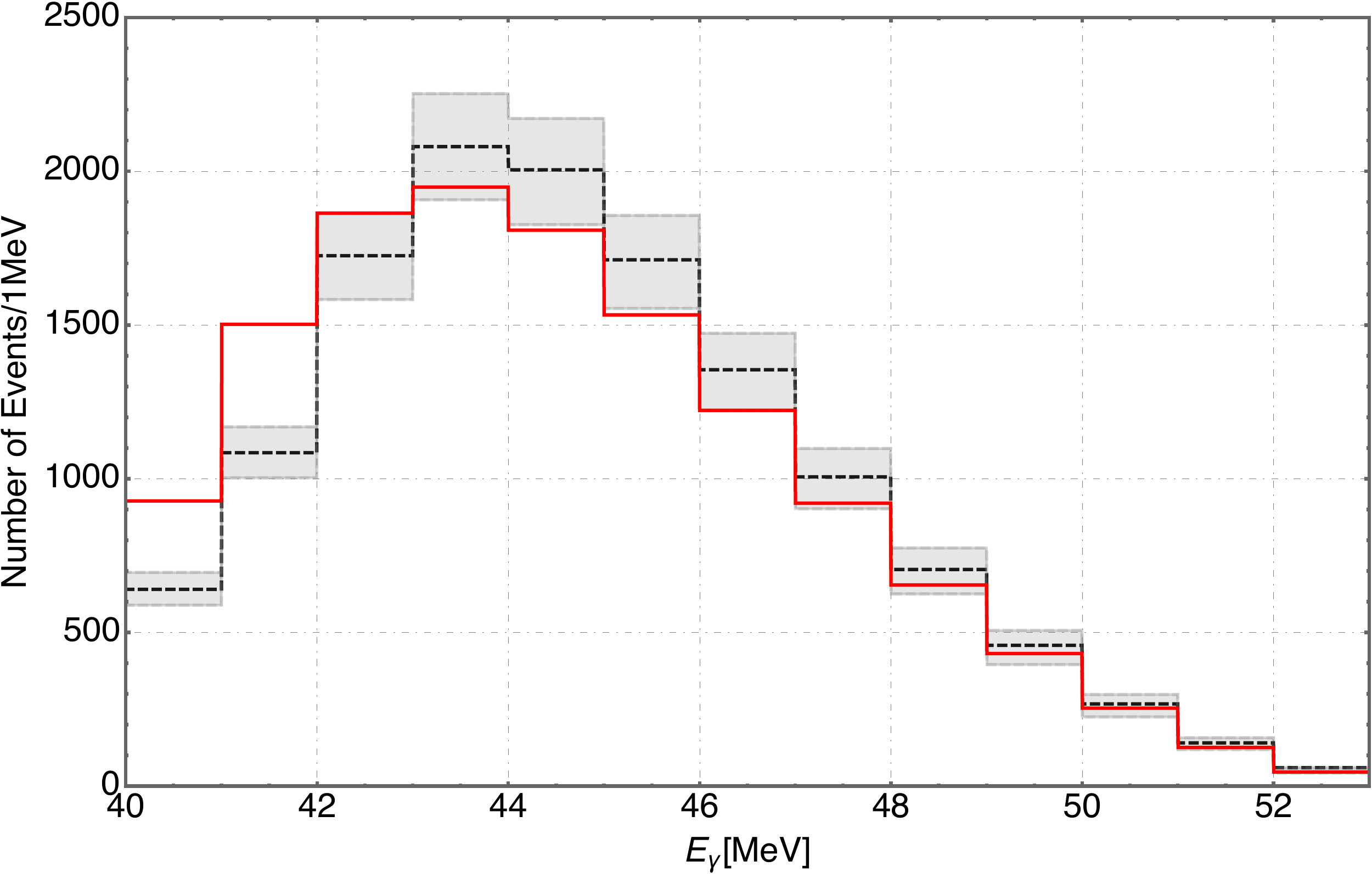}
\includegraphics[width=.32\textwidth]{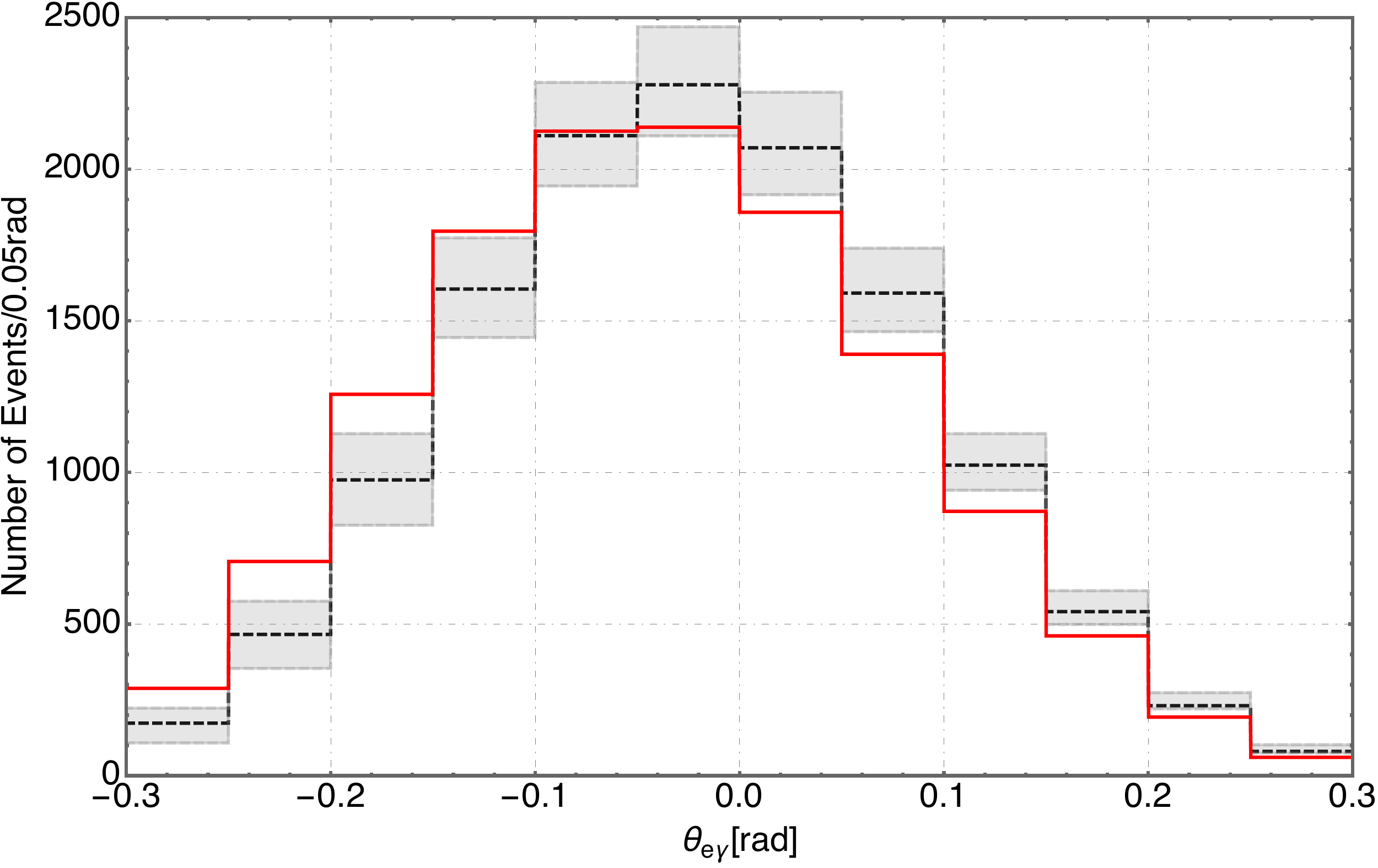}
\caption{Positron energy distribution, photon energy distribution, positron-photon polar angle acollinearity $\theta_{e\gamma}$ distribution  in RMD events using MEG's kinematic cuts and detection efficiencies. Red solid line from our simulation. For a check of validation, we show the central value of the expectation (Black dashed line) and the systematic uncertainty (Gray band) of Monte Carlo simulation given in Ref.~\cite{MEG:2013mmu}.}
\label{fig:MC_validation_RMD}
\end{figure*}

\section{Signal branching ratio}\label{app:signal}

The momentum vectors $\vec p_e$, $\vec p_\gamma$ and $\vec p_a$ always lay in a plane in any frame in which the muon is at rest. We define two such coordinate frames: i) the \textit{polarization frame} ($\theta_{e,\gamma}/\phi_{e,\gamma}$) is the muon rest frame where the $z$-direction is identified with the direction of muon polarization. The orientation of the $x$ and $y$ axis can be chosen arbitrarily. ii) The \textit{positron frame} ($\theta_{e,\gamma}'/\phi_{e,\gamma}'$) is then defined by the Euler rotations
\cite{ParticleDataGroup:2020ssz}
\begin{equation}\label{eq:rotation}
\hat{p}_\gamma (\theta_\gamma,\phi_\gamma) = R_z (\phi_e) \cdot R_y (\theta_e) \cdot  \hat{p}_\gamma' (\theta_\gamma',\phi_\gamma') \ ,
\end{equation}
where $\hat{p}_\gamma$ and $\hat{p}_\gamma'$ are the unit vectors of photon momentum direction in the polarization and positron frame respectively. In this frame the $z$-direction is the direction of positron momentum.  In the evaluation of total branching ratio we use the positron frame, while the angular distributions of signal are most clearly seen in the polarization frame, as the polarization vector correlates with the beam and detector orientation. 

The three body phase space can be written as the integral over the energies of the photon and the positron ($E_\gamma$, $E_e$), the Euler angles of the positron $(\phi_e,\theta_e)$ and the azimuthal angle of the photon around the positron ($\phi'_\gamma$) \cite{ParticleDataGroup:2020ssz}. The total branching ratio of the process $\mu^+ \to e^+ a \gamma$ is therefore given by
\begin{widetext}
\begin{equation}
{\rm BR}(\mu^+ \to e^+ a \gamma)=\!\! \int | \mathcal{M}_{\mu \to e a \gamma}|^2 \frac{dE_e\, dE_\gamma\, d\cos\theta_e \,d\phi_e\, d\phi_\gamma'}{16 (2\pi)^5 m_\mu \Gamma_\mu  }\,, \quad \quad \label{eq:sig_width}
\end{equation}
where $\Gamma_\mu=3\times 10^{-19} \text{ GeV}$ is the total width of muon. The integral runs over the allowed phase space, where a lower cut of $E_\gamma$ is needed to regulate the IR divergence in the matrix element. The squared amplitude is given by
\begin{eqnarray}
| \mathcal{M}_{\mu \to e a \gamma}|^2 & = & \frac{2\pi \alpha_{\rm em} m_\mu^2}{f_a^2 y^2 (1-x-y)} \Bigl [ ( |C_{e\mu}^V|^2 + |C_{e\mu}^A|^2) F_I (x,y) + 2\text{Re}(C_{e\mu}^V {C_{e\mu}^A}^*) P_\mu F_A (x,y,\theta_e, \theta_\gamma) \Bigr ],
\end{eqnarray}
where $P_\mu$ is the muon polarization, which is measured to be \mbox{$P_\mu=-0.85$} at MEG~\cite{MEG:2015kvn}. We further define the functions
\begin{eqnarray}
F_I (x,y,\eta_a) & = & y(1-x^2 - \eta_a^2) - 2(1-\eta_a)(1-x-\eta_a), \\
F_A (x,y,\eta_a,\theta_e, \theta_\gamma) & = & \cos \theta_e \Bigl ( x (2\eta_a + x (2-y)+(\eta_a +1) y -2) \Bigr ) + \cos \theta_\gamma \Bigl ( y (1-\eta_a)(\eta_a+x-1) \Bigr ).
\end{eqnarray}
\end{widetext}
with $x=2E_e/m_\mu$, $y=2E_\gamma/m_\mu$, $\eta_a = m_a^2/m_\mu^2$.
The polarization frame angle $\theta_\gamma$ is a function of $\theta'_\gamma$, $\phi'_\gamma$, $\phi_e$ and $\theta_e$ through the rotation in \eqref{eq:rotation}, which gives
\begin{equation}
\cos \theta_\gamma =\cos\theta_e\cos\theta'_\gamma-\cos\phi'_\gamma\sin\theta_e\sin\theta_\gamma'.
\end{equation}
The positron frame polar angle $\theta_\gamma'$ is in turn just the opening angle between positron and photon, which is fixed for a given value of $E_e$, $E_\gamma$ and $m_a$ as
\begin{eqnarray}
\cos \theta_\gamma' & = & 1 + \frac{2(1-x-y-\eta_a)}{xy}.
\end{eqnarray}

\section{Details of our simulation}\label{app:validation}

In this section we describe the implementation of the different trigger and offline selections in our own Monte Carlo simulation, as well as the detector smearing. The section is structured as follows: in Sec.~\ref{sec:vali} we describe our validation procedure of the Monte Carlo, in Sec.~\ref{sec:res} we discuss the extrapolation of the detector resolution performances beyond the typical signal region of MEG and MEG II. In Sec.~\ref{sec:cuts} we discuss the differential efficiencies of the different search strategies as described in the main text. We also provide further differential distributions of signal and backgrounds. All signal/background events are generated accounting for a muon polarization of $P_\mu = -0.85$, which is the average polarization measured at MEG~\cite{MEG:2015kvn}. 

\subsection{Validation}\label{sec:vali}

Using the approximate trigger efficiency discussed in Sec.~\ref{sec:MEG} and the full differential decay width of RMD process \cite{Fronsdal:1959zzb,Kuno:1999jp}, we validate our simulation by reproducing the distribution of the RMD events as a function of $E_e$, $E_\gamma$ and $\theta_{e\gamma}$ separately. This is shown in  Fig.~\ref{fig:MC_validation_RMD}. 

Except for two lowest photon energy bins, the distributions generated with our Monte Carlo reproduce the MEG distributions Ref.~\cite{MEG:2013mmu} quite well, within their systematic uncertainties. The biggest deviations are at low $E_\gamma$, where our extrapolation of the trigger efficiency is expected to fail. 

For the MEGII-RMD analysis we use the same set of events as for the MEG-RMD case. For the MEGII-ALP analysis, signal/background event sets are obtained using the same procedure, but with the different kinematic selection explained in Sec.~\ref{sec:MEGII}. As this is a projection for a future search, we do not have a way of validating it with existing data. 

\begin{table*}[t]
\centering
\begin{tabular}{c|c|c|c|c|c|c|c|c|c|c}
& & \multicolumn{9}{c}{Efficiencies} \\
\cline{3-11}
 & $\text{BR}_i^{\text{base}}$ & \multicolumn{3}{c|}{$\epsilon_i^{\text{trig.}}$} & \multicolumn{5}{c|}{$\epsilon_i^{\text{off.}}/\epsilon_i^{\text{trig.}}$} & ALP search \\
\cline{3-11}
& & $\epsilon_{E_e}$  & $\epsilon_{E_\gamma}$ & $\epsilon_{\theta_{e\gamma}}$ & timing & $E_\gamma > 45$ MeV & $E_\gamma > 40$ MeV & $|\theta_{e\gamma}|<0.3$ & $|\phi_{e\gamma}|<0.3$ & $\Delta m_{\slashed{E}}^2$ \ \\
\hline
$\mathcal{B}_{\text{RMD}}$ & $1.44 \times 10^{-5}$ & $0.15$ & $5.3 \times 10^{-4}$ & $0.49$ & 1.0 & $0.83$ & $0.90$ & $0.99$ & $0.48$ & $0.21$ \\
\hline
$\mathcal{B}_{\text{RC}}$ & $7.08 \times 10^{-4}$ & $0.34$ & $0.01$ & $0.03$ & $0.17$ & $0.98$ & $0.93$ & $0.92$ & $0.13$ & $0.22$ \\
\hline
$\mathcal{S}$ & $4.8 \times 10^{-9}$ & $0.39$ & $1.68 \times 10^{-3}$ & $0.48$ & 1.0 & $0.93$ & $0.94$ & $0.98$ & $0.46$ & $0.74$ \\
\end{tabular}
\caption{Efficiencies in each kinematic selection for the parasitic analysis (Sec.~\ref{sec:MEGlimit}). For signal, $F_{\mu e}^{V-A}$ is fixed to $10^9$ GeV. The ``timing'' selection refers to the tighter coincidence requirement in the offline cuts, as compared to the trigger selection. See Sec.~\ref{sec:trigger} for details.}
\label{table:efficiency_MEGRMD}
\end{table*}

\begin{table*}[t]
\centering
\begin{tabular}{c|c|c|c|c|c}
& & \multicolumn{4}{c}{Efficiencies} \\
\cline{3-6}
& $\text{BR}_i^{\text{base}}$ & \multicolumn{2}{c|}{$\epsilon_i^{\text{trig.}}$} & \multicolumn{2}{c}{$\epsilon_i^{\text{off.}}/\epsilon_i^{\text{trig.}}$} \\
\cline{3-6}
& & \ $\epsilon_{E_e}$ \ & \ $E_\gamma > 10$ MeV \ & \ timing  \ & \ $\Delta m_{\slashed{E}}^2$  \ \\
\hline
$\mathcal{B}_{\text{RMD}}$ & $1.44 \times 10^{-5}$ & 0.15 & 0.29 & 1.0& 0.08 \\
\hline
$\mathcal{B}_{\text{RC}}$ & $7.08 \times 10^{-4}$ & 0.34 & 0.10 & 0.17 & 0.04 \\
\hline
$\mathcal{S}$ & $4.8 \times 10^{-9}$ & 0.39 & 0.36 & 1.0 & 0.81 \\
\end{tabular}
\caption{Efficiencies in each kinematic selection for the dedicated run (Sec.~\ref{sec:MEGII}). For signal, $F_{\mu e}^{V-A}$ is fixed to $10^9$ GeV. The ``timing'' selection refers to the tighter coincidence requirement in the offline cuts, as compared to the trigger selection. See Sec.~\ref{sec:trigger} for details.}
\label{table:efficiency_MEGIIALP}
\end{table*}

\subsection{Detector resolution}\label{sec:res}
The MEG detector resolutions for positron and photon energies between 40 and 53 MeV is extracted from Ref.~\cite{MEG:2020zxk} fitted and extrapolated to a wider energy range of energies with the following functional dependencies:
\begin{eqnarray}
\frac{\delta E_\gamma^{\text{MEG}}}{E_\gamma} & = & \Bigl [ 2.0\% \oplus \frac{1.18\%}{\sqrt{E_\gamma/\text{MeV}}} \Bigr ], \\
\frac{\delta E_{e^+}^{\text{MEG}}}{E_{e^+}} & = & \Bigl [ \frac{0.46\%}{(E_{e^+}/45\text{MeV})^2} \oplus \frac{0.78\%}{(E_{e^+}/45\text{MeV})^{9/4}} \Bigr ], \\
\delta \theta_{e^+/\gamma}^{\text{MEG}} & = & \Bigl [ \frac{0.126}{(E/45\text{MeV})^{7/4}} \oplus \frac{0.153}{(E/45\text{MeV})^2} \Bigr ] \text{ mrad}, \quad\quad \\
\delta \phi_{e^+/\gamma}^{\text{MEG}} & = & \Bigl [ \frac{0.31}{(E/45\text{MeV})^{9/4}} \Bigr ] \text{ mrad}\, . 
\end{eqnarray}
The  functional form of the photon energy resolution is the typical form for any calorimeter~\cite{Fabjan:2003aq}, where the stochastic term drops as $1/\sqrt{E}$ and the constant term accounts for effects that are independent on the particle energy.
We use the fits above in our Monte Carlo to compute the smearing of the energy, angle and missing invariant mass distributions at MEG. 

For MEG II we take into account the improved resolutions of the detector with respect to MEG. In practice, we replace the MEG resolutions at $E_{e^+/\gamma} \simeq m_\mu/2$ with the ones provided in Table~8 of Ref.~\cite{MEGII:2018kmf} 
\begin{eqnarray}
\delta E_{e^+}^{\rm MEG II} & = & 0.34 \delta E_{e^+}^{\rm MEG},\\
\delta E_\gamma^{\rm MEG II} & = & 0.51 \delta E_\gamma^{\rm MEG},\\
\delta \theta_{e^+/\gamma}^{\rm MEG II} & = & 0.56  \delta \theta_{e^+/\gamma}^{\rm MEG},\\
\delta \phi_{e^+/\gamma}^{\rm MEG II} & = & 0.43\delta \phi_{e^+/\gamma}^{\rm MEG}\ .
\end{eqnarray}
We then extrapolate the MEG II resolutions to lower energies by using the same functional dependence as the one derived for MEG. In this way the resolution improvement of MEG II with respect to MEG is essentially an overall rescaling of the resolution, independent on energy. This assumption should be revisited once the performance information of the MEG II detector at lower energies is available.

\subsection{Online and offline efficiencies}\label{sec:cuts}
In this section we summarize the cut flow for the data taking strategies discussed in this paper:  i) the parasitic analysis of the MEG RMD data presented in Sec.~\ref{sec:MEGlimit} and its projection at MEG II showed in Sec.~\ref{sec:MEGIIlimit}, and ii) the MEG II-ALP dedicated run discussed in Sec.~\ref{sec:MEGII}. In Table~\ref{table:efficiency_MEGRMD} we give the  integrated efficiencies of both the trigger and the offline selection for the MEG RMD data taking for the ALP signal and the RMD and the RC backgrounds. The efficiencies are normalized with respect to baseline branch ratios defined in Sec.~\ref{sec:trigger}. The baseline cuts are $E_{e}>40\text{ MeV}$, $E_{\gamma}>5\text{ MeV}$, $\theta_\gamma \in [70^\circ, 110^\circ]$ and $\phi_\gamma \in [-60^\circ, +60^\circ]$, which account for the geometric acceptance of the MEG photon detector. We moreover impose $\phi_e \in [120^\circ, 240^\circ]$, to ensure that the positron is approximately within the acceptance of the timing detector. Every efficiency is normalized with respect to the number of events passing the previous cut, from left to right, such that the product of all the trigger requirements (columns 3, 4 and 5) is reproducing the total trigger efficiency discussed in Sec.~\ref{sec:MEG}. The numbers in the $\epsilon_i^{\text{off.}}/\epsilon_i^{\text{trig.}}$ column indicate the sequential loss in efficiency once the offline selections are imposed, relative to the trigger selection. In other words, the total offline efficiency can be obtained by multiplying the numbers in columns 3 to 8.

\begin{figure*}[t]
\centering
\includegraphics[width=\textwidth]{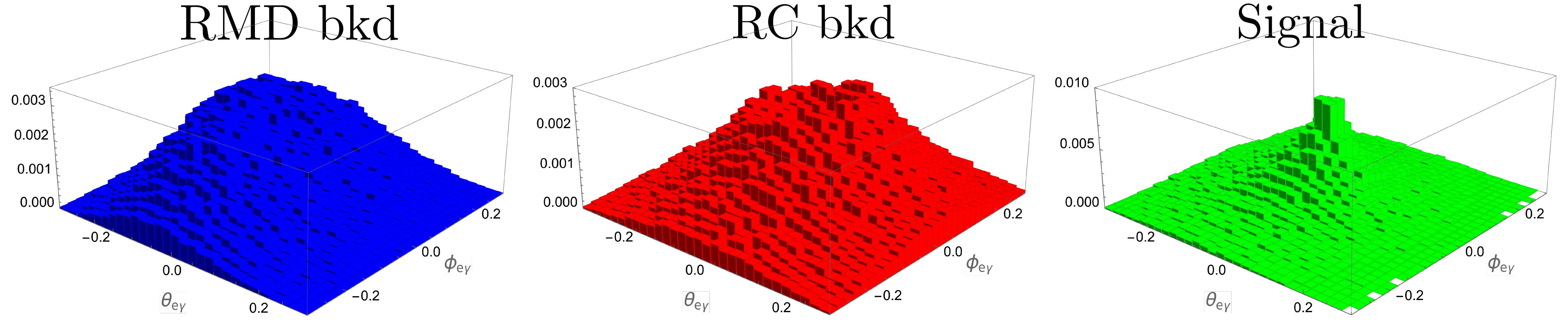}
\caption{The unit-normalized distributions of RMD (Left), RC (Middle) background and signal (Right) events on the ($\theta_{e\gamma}, \phi_{e\gamma}$)-plane for the parasitic analysis ({\color{Green} $\bullet$} in Fig.~\ref{fig:money}) with the bin size of 20mrad ($\theta_{e\gamma}$) $\times$ 20mrad ($\phi_{e\gamma}$). For the signal we took $m_a = 10^{-4}$ MeV.}
\label{fig:Theta_egamma_dist}
\end{figure*}

The table shows that the trigger requirement on the photon energy in Fig.~\ref{Fig:MEGrec}, together with the cut on the positron energy, is the main limitation on MEG sensitivity for ALPs, while the angular cut is an $\mathcal{O}(1)$ effect in this ordering.\footnote{Note that this statement depends on the ordering of the cut flow. We checked that taking any pair out of the three trigger requirements in Fig.~\ref{Fig:MEGrec},  would select a back-to-back topology for the final states leading to a similar suppression of the signal compared to the background.} This trigger selection has essentialy two main drawbacks: i) the small signal efficiency ii) the background shape of the RC background, which becomes very similar to the signal after the trigger requirements are imposed. This second issue makes the offline variables quite inefficient in separating the signal from the background, as can be seen directly from the signal and background distributions in the left-hand panel of Fig.~\ref{fig:MMdist} as well as Fig.~\ref{fig:Theta_egamma_dist}.

\begin{figure*}[t]
\centering
\includegraphics[width=\textwidth]{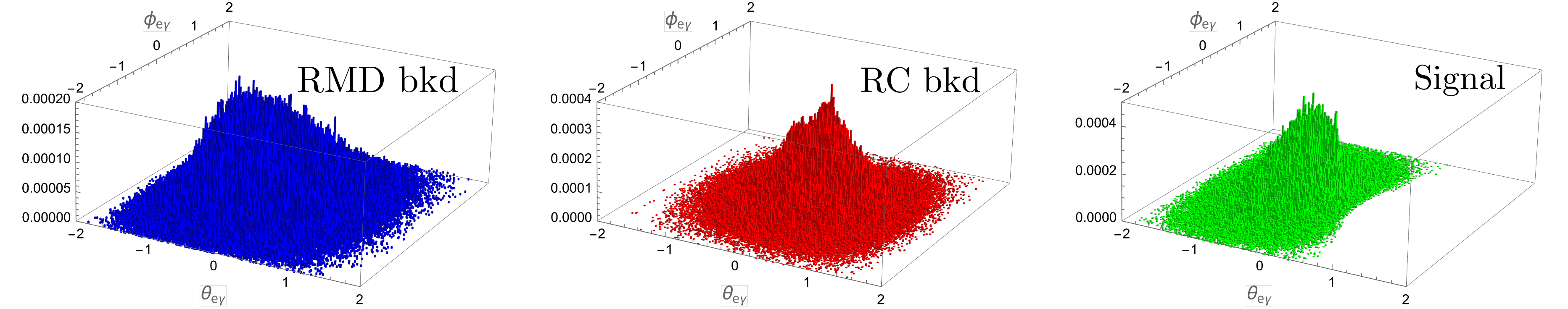}
\caption{The unit-normalized distributions of RMD (Left), RC (Middle) background and signal (Right) events on the ($\theta_{e\gamma}, \phi_{e\gamma}$)-plane for the benchmark point of the dedicated run ({\color{Purple} $\star$} in Fig.~\ref{fig:money}) with the bin size of 20mrad ($\theta_{e\gamma}$) $\times$ 20mrad ($\phi_{e\gamma}$) up to modulo of 2$\pi$ for $\phi_{e\gamma}$. For the signal we took $m_a = 10^{-4}$ MeV.}
\label{fig:Theta_egamma_dist2}
\end{figure*}

In Table \ref{table:efficiency_MEGIIALP} we show the integrated efficiencies of the MEG II-ALP data taking strategy.  Reducing the beam intensity allows on the one hand to increase the signal efficiency at trigger level and on the other hand to keep the shape of the RC flat enough to be more easily distinguishable from the signal shape in the offline analysis. This can be seen in the missing mass distribution in the right-hand panel of Fig.~\ref{fig:MMdist}, where the RC background appears as a featureless flat distribution, and from the angular distributions of Fig.~\ref{fig:Theta_egamma_dist2}. 

\subsection{Angular differential distributions}\label{sec:diffangle}
For completeness we show the angular distributions of the ALP signal and the RMD and RC backgrounds for the parasitic analysis of the MEG RMD data in Fig.~\ref{fig:Theta_egamma_dist} and for the MEG II-ALP dedicated run in Fig.~\ref{fig:Theta_egamma_dist2}. These are events passing both the trigger and the offline selection, where we applied the cut on the missing mass window. By comparing the two figures it is clear that the standard MEG RMD trigger selection produces a very different shape for the RC background than with the  MEG II-ALP selection. While difficult to see by eye, a likelihood analysis reveals that the MEG II-ALP selection yields a significantly better signal vs background separation. 

\section{Different chirality structures}\label{app:othercases}

\begin{figure}[h]
\centering
\includegraphics[width=\textwidth]{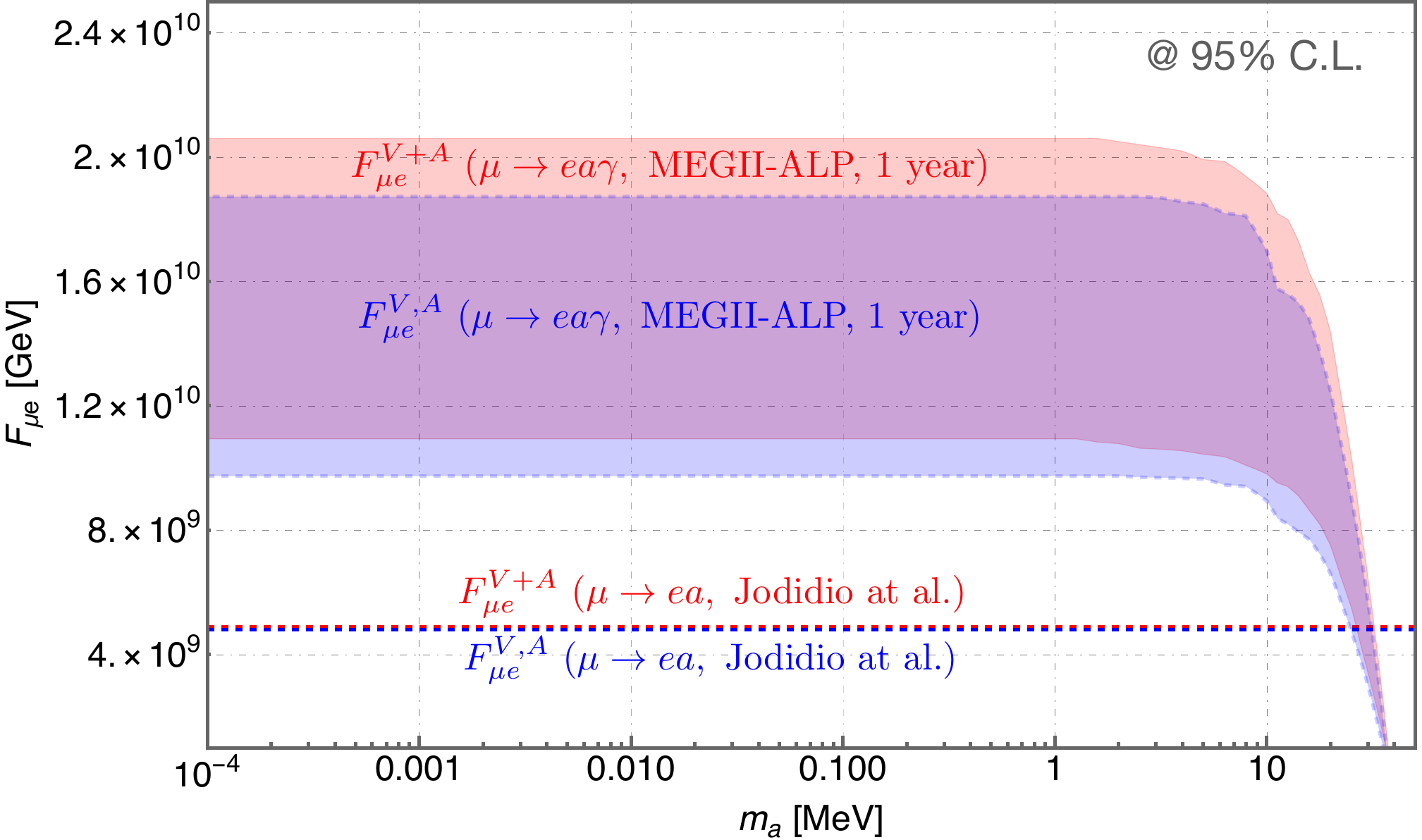}
\caption{95\% C.L. limits on $F_{\mu e}$ for different chiralities. {\bf Red (blue) solid} bands indicates the reach of $\mu \to ea\gamma$ search with $V+A$ ($V$ or $A$) interaction for LFV coupling of ALP, assuming 1 year of running time. The width of the bands correspond to the uncertainty in our estimation of the trigger rate as shown in Fig.~\ref{fig:trigger}. {\bf Dashed} lines are the current limits from the $\mu^+ \to e^+a$ search performed by Jodidio et al. in 1986 \cite{Jodidio:1986mz} for comparison.}
\label{fig:llimit_all}
\end{figure}
We show here the reach of our dedicated data taking proposal for different chiral structures of the axion couplings to leptons. These are shown in Fig~\ref{fig:llimit_all}. Interestingly, even for the most conservative estimate of our expect trigger rate, the expected sensitivity of MEG II with our data taking proposal and 1 year of data taking can surpass the current best limit coming from the experiment of Jodidio et al.~\cite{Jodidio:1986mz} for right-handed ALP couplings (V+A) or purely axial (purely vectorial) couplings. The reach in these scenarios is sensibly improved compared to the V-A case discussed in the main text due to the more distinctive angular distribution of the signal events with respect to the background events. 

\bibliographystyle{JHEP}
\bibliography{FV_lepton}

\end{document}